\documentclass[aps,pre,floatfix,twocolumn,superscriptaddress]{revtex4-2}
\usepackage{graphicx}
\usepackage{float}
\usepackage{xcolor}
\usepackage{commath,amssymb,amsmath,stmaryrd,tabularx}

\DeclareMathOperator{\csch}{csch}

\renewcommand{\thesubsection}{\thesection.\arabic{subsection}}

\makeatletter
\renewcommand{\p@subsection}{}
\renewcommand{\p@subsubsection}{}
\makeatother

\begin{document}
\title{Hierarchical Loop Stabilization in Periodically Driven Elastic Networks}
\author{Purba Chatterjee}
\affiliation{Department of Physics and Astronomy, University of Pennsylvania, Philadelphia, PA $19104$, USA}
\author{Eleni Katifori}
\affiliation{Department of Physics and Astronomy, University of Pennsylvania, Philadelphia, PA $19104$, USA}
\affiliation{Center for Computational Biology, Flatiron Institute, New York, NY $10010$, USA}

\begin{abstract}
Network remodeling, or adaptation, in the presence of periodically driven forcings has hereto remained largely unexplored, despite the fact that a broad class of biological transport networks, e.g. animal vasculature, depends on periodic driving (pulsatility of the heart) to maintain flow. Here we propose a new adaptation model that incorporates the effect of short-term flow pulsatility into the long-term remodeling signal for periodically driven elastic networks, and is consistent with optimality principles of dissipation and metabolic cost minimization. Using this model to adapt hierarchical elastic networks with multiple levels of looping, we show that very different network architectures are possible at steady-state depending on the driving frequency of the pulsatile source and the geometric asymmetry of the paths between the externally driven nodes of the network. Specifically resonant frequencies are shown to prioritize the stabilization of fully looped structures or higher level loops proximal to the source, whereas anti-resonant frequencies predominantly stabilize loop-less structures or lower-level loops distal to the source. Thus, this model offers a mechanism that can explain the stabilization of phenotypically diverse loopy network architectures in response to source pulsatility under physiologically relevant conditions and in the absence of other known loop stabilization mechanisms, such as random fluctuations in the load or perfusion homogenization. 
\end{abstract}

\maketitle
\section{Introduction}\label{secIntro}
Biological fluid transport networks like animal vasculature or the protoplasmic veins of the slime mould \textit{Physarum polycephalum} are crucial building blocks of complex life. These networks are predominantly composed of vessels that are compliant (i.e. elastic) and can change their shape locally to accommodate changes in the flow. Their structure is thought to be optimized for the crucial functions they perform for the organism, which includes but is not limited to the distribution of oxygen and nutrients, the removal of metabolic waste products, and long range transmission of information. The structural phenotype of biological transport networks range from large-scale tree-like architectures, like the arteriovenous tree in mammals that serves entire organs, to highly reticulated or ``loopy" structures, like the microvasculature (capillary network) that caters to individual organs \cite{Nelson1997,laguna2008,ronellenfitsch2019}. Loops are ubiquitous in biological networks, and are understood to be beneficial to the network given their role in mitigating vessel-damage and facilitating robustness to source fluctuations \cite{katifori2010,corson2010,kaiser2020}. Given the importance of the structure of these networks for biological function, adaptation, or the remodeling of network elements in response to changing environmental signals, is therefore an important and widely studied phenomenon.\cite{pries1998,pries2008,hu2013,ronellenfitsch2016,chang2019,gounaris2024central,kramer2023,marbach2023}

Many transport networks depend on pulsatile sources to maintain flow. The most prominent example of this is animal vasculature, where the flow has pulsatile components imparted by the periodic beating of a centralized power source, the heart \cite{travasso2025predicting}. The presence of pulsatility in the flow necessarily increases the net energy dissipation across the network, in comparison to steady flow with the same average flow rate. The amplitude of pulsatility is known to decrease with increasing distance from the source, owing to fluid inertia and the compliance of the elastic vessel walls. Yet, despite this damping effect, pulsatility is never completely dispensed with and exists to some extent even at the level of capillaries \cite{harazny2014}. Previous work has shown that the energetic advantage of decreasing flow pulsatility by increasing vessel compliance, comes at the cost of increased response times to sudden changes in source current or pressure waveforms, for example a change in the heart rate \cite{fancher2022,fancher2023tradeoff}. In fact, experimental evidence shows that both an increase or decrease in arterial pulsatility can lead to detrimental pathologies in mammals \cite{cheng2014,bartoli2010,muhire2019}, suggesting that maintaining the right levels of pulsatility in the flow is crucial for biological function.  

Despite the important role played by pulsatility in biological flows, adaptation in compliant vessel networks driven by pulsatile sources has remained largely unexplored thus far, although fluctuating sources and sinks have been implemented in previous theories \cite{katifori2010,hu2013,grawer2015,waszkiewicz2024,gyllingberg2025minimal}. Most existing models of adaptation assume that vessels respond to the flow in a non-dynamical manner on short timescales: the flow relaxes instantaneously to the average value at the source at every point along the length of the vessel. In reality however, changes in the flow are opposed by the finite inertial resistance of the fluid and adjusted for locally in elastic vessels by affecting a change in their shape. The combination of these effects yields a short timescale of flow relaxation dynamics which results in time and length averaged flow rates in elastic vessels that differ from the mean source current, unlike in the non-dynamical approximation \cite{fancher2022}.  

Previous work has shown that when the short-term pulsatile dynamics of flow in compliant vessels is accounted for in the driving signal of a widely used remodeling rule for vascular development, it yields long-time network structures that are fundamentally different from those expected when vessels can only respond non-dynamically to the flow at short timescales \cite{chatterjee2024}. Specifically, looped architectures can be stabilized for a much broader class of physiologically relevant metabolic constraints than predicted by existing theories, mediated by resonances that amplify the energy dissipation in individual vessels. This finding further confirms the indispensability of pulsatility despite the energy cost that accompanies it, and offers a possible explanation for the observed ubiquity of loops in biological networks, especially closer to the pump, where the pulsatility is larger and perfusion considerations are less important.

However, most remodeling rules, including the one considered in \cite{chatterjee2024}, consider driving signals that are appropriate for steady flow sources, necessitating a modified approach to account for networks driven by a pulsatile source. Moreover, while short-term pulsatility can be shown to facilitate loop stabilization \cite{chatterjee2024}, what remains to be investigated is whether this same mechanism can account for the broad spectrum of structural phenotypes from tree-like to reticulated network architectures. 

In this paper, we attempt to address these important questions in an effort to formulate a  mechanics informed theory of adaptation in the presence of pulsatility that can better estimate the long-time structures of complex biological networks, especially in the absence of other loop stabilizing factors. Motivated by other work that established gradient descent based adaptation rules for biological flow networks \cite{hu2013}, we start from the standard principles of energy optimization and derive an adaptation rule for periodically driven elastic flow networks that addresses the effect of short-term pulsatile flow dynamics on the long-term remodeling signal at the timescale of adaptation. Looking at progressively more hierarchical networks we show that short-term pulsatility can stabilize very different network architectures depending on the frequency of periodic driving and the length asymmetry of the different paths between the externally driven nodes of the network. Resonant frequencies, which amplify the flow in a vessel, are shown to prioritize loops proximal to the source (higher-level loops) for stabilization, yielding reticulated structures, whereas anti-resonant frequencies lead to shunting (loop destabilization) or stabilize loops distal to the source (lower-level loops), leading to more tree-like structures. 

Moreover, our results appear to be largely independent of the detailed form of the remodeling rule used  to evolve the networks. This suggests that rather than the exact adaptation equation used to evolve networks, it is indeed the short-term pulsatile dynamics in individual elastic vessels, or in other words, the ability to adapt in a compliant manner to the flow on short-time-scales, that is responsible for robust loop stabilization. While our simplified flow dynamics model and toy networks cannot possibly capture the vast complexity of actual animal vasculature, it nevertheless constitutes an important theoretical step in the direction of better understanding the interplay between the short and long timescales of network remodeling in the presence of pulsatile driving, and its possible role in stabilizing a broad array of structural phenotypes in adapting biological networks.

\section{The Adaptation Model}\label{sec:model}
Network adaptation or remodeling is typically studied within the framework of optimization, i.e. evolving the network to minimize the net energy dissipation while satisfying constraints imposed by material or metabolic costs. A remodeling rule that has been used extensively to study the evolution of vessel radii in plant and 
animal vascular development, as well as for the foraging canals of the slime mold \textit{Physarum polycephalum} \cite{tero2010,marbach2023,hacking1996,rollandLagan2005,hu2013,berkel2013,ronellenfitsch2016,ronellenfitsch2019} is given by
\begin{equation}\label{eq:AE_S}
    \frac{dR_{e}}{dt'}=\frac{a\langle Q_{e}^2\rangle^\gamma}{R_{e}^3}-bR_{e},
\end{equation}
where $R_e$ and $Q_e$ are the radius and the current in the $e^{th}$ vessel of the network respectively, $\gamma \in (0,1)$ characterizes a metabolic cost, and $a$ and $b$ are constants. The steady states of this adaptation rule are identical to those of an energy functional of the form 
\begin{equation}\label{eq:EF_S}
    E_s=\sum_e\left(\frac{\langle Q_e^2\rangle}{K_e}+C_0 K_e^{1/\gamma-1}\right)L_e,
\end{equation}
where $C_0$ is a constant, $L_e$ is the length of vessel $e$ and $K_e$ is the conductance of the $e^{th}$ vessel, which for Poiseuille flow relates to the radius as $K_e\propto R_e^4$.
The first term in the summation is the power dissipated in the vessel in the case of steady flows, $D_e=\langle Q_e^2\rangle L_e/K_e$ (or $D_e=\langle Q_e^2\rangle L_e/R_e^4$ for Poiseuille flow). The second term captures a class of metabolic or material costs which have a power-law dependence on the vessel conductance, and which competes with energy minimization during remodeling \cite{hu2013,ronellenfitsch2016}. For example, in animal vasculature, a blood volume cost would translate to a $\gamma=2/3$ exponent assuming Poiseuille flow, consistent with Murray's law \cite{murray1926physiological}. A vessel surface area cost would lead to $\gamma=4/5$. Continuously varying $\gamma$ in the range $(0,1)$ thus allows us to study a broad class of costs that could affect the long-time structure of the network, even if only certain discrete values of $\gamma$ correspond to realistic metabolic or material costs.

Note that Eq.~\ref{eq:AE_S} can be rewritten in terms of the conductance $K_e$ of the vessel. The growth term (first term) in Eq.~\ref{eq:AE_S} is counterbalanced by the decay term (second term) which prevents the radius of each vessel from growing without bounds. The growth term of the adaptation rule carries the driving signal for remodeling - the vessel mean-squared current averaged over the length of the vessel. Eq.~\ref{eq:AE_S} is typically used for remodeling with steady flows, for which the mean-squared current in a vessel is given by the mean-squared source current split proportionally among each vessel in accordance with their radii. Increasing $\gamma$ increases the influence of the growth term on the adaptation dynamics.

\subsection{Remodeling Rule for Pulsatile Flows}\label{subsec:AE}
There is an abundance of experimental support for the hypothesis that the structure of biological networks is governed by optimality principles based on energy and cost minimization \cite{sherman1981connecting, kassab1995pattern,zamir1979arterial, kassab2006scaling,huo2009scaling,taylor2024systematic,zhou1999design,kamiya1972optimal,mayrovitz1983microvascular,taber2001investigating,hughes2015optimality,silva2015scaling,painter2006pulsatile,taylor1967elastic,shumal2025association,hahn2008comparison,schoenenberger2012deviation,hughes2015optimality}. The predominance of the optimization framework in the literature on structural adaptation \cite{kamiya1984adaptive, chen2012haemodynamics,banavar2000topology,durand2006architecture,hu2013,ronellenfitsch2019,ronellenfitsch2016,marbach2023,tero2010,hacking1996,rollandLagan2005,berkel2013,corson2010,bohn2007structure} is moreover a reflection of its robustness as a means of studying long-term network remodeling on the basis of physical and mechanistic considerations, even when the putative optimum can only approximately be approached by biology. Motivated by this consideration, we now seek to construct an energy functional similar to Eq.~\ref{eq:EF_S}, but for networks driven by a pulsatile source. It is generally accepted in the well developed theory of structural adaptation in biological networks that radii evolution is driven by the vessel response to the wall shear stress $\sigma$. While the vessel averaged wall shear-stress by itself has been shown to be inadequate to control the growth of vascular networks \cite{kurz2000,hu2012,hacking1996,ronellenfitsch2016,ronellenfitsch2019,eichmann2005,scianna2013,qi2024}, a realistic driving signal for adaptation must however have an explicit functional dependence on $\sigma$. With pulsatile drivers, the power dissipated in a vessel $e$ is given by $D_e=\sigma_e Q_e L_e/R_e$ \cite{zamir2002} (see Sec.~A2 of the appendix). Assuming Poiseuille flow, the energy functional for periodically driven elastic networks can be written as
\begin{equation}\label{eq:EF_P}
    E_p=\sum_e\left(\frac{\langle \sigma_e Q_e\rangle}{R_e}+C_0 R_e^{4(1/\gamma-1)}\right)L_e,
\end{equation}
where the average of $\sigma_e Q_e$ is now evaluated over one cycle of periodicity, as well as over the length of the vessel:
\begin{equation}\label{eq:sigmaQ}
\langle \sigma_e Q_e \rangle= \frac{1}{T}\int_0^T dt \hspace{5pt}\left[\frac{1}{L_{e}}\int_0^{L_{e}} dz \hspace{5pt} \sigma_e(z,t) Q_e(z,t)\right].
\end{equation}
Here $T=2\pi/\omega$ is the well-defined time period of the pulsatile source. Note that the timescale $t$ of flow relaxation dynamics in individual vessels of the network is much smaller than the adaptation timescale $t'$. It is precisely this timescale separation that allows us to assume that the network has encountered several pulsatility cycles before any infinitesimal change in conductance due to remodeling takes place, resulting in a stable average flow state which can drive radii evolution on the timescale of adaptation. We assume that the radius is constant over the length of the vessel and on timescales shorter than $t'$ and therefore does not require averaging. 

Eq.~\ref{eq:EF_P} has minima that can be approached by an adaptation rule of the form
\begin{equation}\label{eq:AE_P}
    \frac{dR_{e}}{dt'}=\frac{a\langle \sigma_e Q_e\rangle^\gamma}{R_{e}^{3(1-\gamma)}}-bR_{e}.
\end{equation}
The well separated time-scales of short-term dynamics and long-term adaptation guarantees that the effect of the short time-scale is contained entirely in $\langle \sigma_e Q_e\rangle$ in Eq.~\ref{eq:AE_P}, making a direct quantitative comparison with the adaptation time-scale ($t'$ and $b^{-1}$, the decay time) unnecessary.
For steady flows, the wall shear stress has a simple relation to the flow in a vessel, given by $\sigma\propto Q/R^3$, which reduces Eq.~\ref{eq:EF_P} to the steady flow energy optimization functional given in Eq.~\ref{eq:EF_S}, and the adaptation rule for pulsatile flows in Eq.~\ref{eq:AE_P} to the usual steady flow form given in Eq.~\ref{eq:AE_S}.

Eq.~\ref{eq:AE_P} must be seen as a self-consistent phenomenological model for long-term remodeling dynamics in periodically driven networks, within the framework of optimization. It is agnostic to what physical quantities the vessels comprising the network actually locally sense and respond to, about which there is a lack of general consensus beyond the accepted view that they must depend on the wall shear stress experienced by vessels. We have chosen this approach because our focus in this work is not on the exact form of the remodeling rule, but on establishing the importance of the interplay between the dynamical response of elastic vessels to flow-pulsatility on short timescales, and the long-term structures of these adapting networks. Fortunately, as we will show, different adaptation rules, and even ones not derived from an energy functional, generate qualitatively identical and quantitatively similar network structures as long as the driving signal takes into account the short-term flow pulsatility in vessels (see Sec.~A4 of the appendix).
\subsection{Flow dynamics Model for Compliant Vessels}\label{subsec:CompliantVessels}
Following \cite{fancher2022,chatterjee2024}, we adopt a simplified flow dynamics model in which each compliant vessel is considered to be a thin cylinder whose length $L$ is much larger than its radius $R$. Starting from the Navier-Stokes equation for incompressible flows, and making the standard simplifying assumptions of rotational symmetry and parabolic velocity profiles, the flow volume and momentum conservation equations for a fluid of viscosity $\mu$ and density $\rho$ can be recast into a set of coupled differential equations for the axial current $Q(z,t)$ and pressure $P(z,t)$ at every point along the length of a vessel, as
\begin{align}
\frac{\partial Q}{\partial z} &+c\frac{\partial P}{\partial t}=0,\label{eq:PQ1}\\
\frac{\partial P}{\partial z} + &l\frac{\partial Q}{\partial t} +rQ=0,\label{eq:PQ2}
\end{align}
where $l=\rho/A$ and $r=8\pi\mu/A^2$ are the fluid inertia and flow resistance per unit length in a vessel of cross-sectional area $A=\pi R^2$ (see Sec.~A1 of the appendix for derivation). The vessel compliance $c=\partial A/\partial P$ is defined assuming that the cross-sectional area scales linearly with the fluid pressure, and that the changes in cross-sectional area in response to changes in pressure are sufficiently small at short time-scales. This has the simplifying effect of keeping the compliance $c$ approximately fixed across the length of the vessel and on short time-scales, while allowing it to vary over the timescale of adaptation as the radius $R$ (and thus $A$) evolves, keeping the distensibility fixed such that $c\propto A$ (see Sec.~A1 of the appendix). A characteristic length-scale $\lambda$, a characteristic time-scale $\tau$ and a characteristic admittance scale $\alpha$, each scaling proportional to the area of cross-section $A$, can be derived from the network parameters $c$, $l$ and $r$,
\begin{align}\label{eq:lamtau}
\lambda&=\lambda_0 (R/R_0)^2=\frac{2}{r}\sqrt{\frac{l}{c}},\nonumber\\
\tau&=\tau_0 (R/R_0)^2=\frac{2l}{r},\nonumber\\
\alpha&=\alpha_0 (R/R_0)^2=\sqrt{\frac{c}{l}},
\end{align}
where $R_0$ is a typical radius. The constants $\lambda_0$, $\tau_0$ and $\alpha_0$ allow us to isolate the radius dependence of these characteristic scales, and are set to the same value for each vessel in the network while their radii can evolve independently of one another on the time-scale of adaptation. The time-scale $\tau$ captures the ratio of inertial and viscous forces and approximates the time taken for local flow relaxation in a vessel of radius $R$. The parameter $\tau_0$ can thus be used to characterize the short time-scale of the problem: the limit of the re-scaled frequency $\omega\tau_0$ going to zero recovers the non-dynamical approximation which most adaptation models operate on. In other words, as $\omega\tau_0\to0$, the short-time scale vanishes for a vessel of typical radius, i.e. the flow in it relaxes instantaneously to the average value at the boundary along the entire length of the vessel.  

The length-scale $\lambda$ characterizes the distance over which pulsatility is damped in a vessel of radius $R$. For the typical radius $R_0$, the pulsatile components of the flow are effectively attenuated in vessels with $L/\lambda_0>>1$. As the radius of the vessel decreases, the damping distance $\lambda$ for a vessel becomes shorter, rendering the vessel less able to sustain flow pulsatility. Thus even with finite compliance, a vessel may respond to periodic driving in a non-dynamical manner if the damping length-scale is much shorter than its length. The two parameters $\tau_0$ and $\lambda_0$ allow us to qualify the response of a network with many vessels of typical radius but differing lengths, to pulsatile driving at a given well-defined frequency $\omega$.

 Decomposing $Q(z,t)$ and $\sigma(z,t)$ into their Fourier modes, the relation between the axial current and the wall shear stress is given by \cite{zamir2002}
 \begin{align}\label{eq:sigQrel}
     Q(z,t)&=\sum_{n=-\infty}^{n=\infty}\tilde{Q}^{(n)}(z) e^{in\omega t},\nonumber\\
     \sigma(z,t)&=\sum_{n=-\infty}^{n=\infty}\tilde{f}^{(n)}\tilde{Q}^{(n)}(z) e^{in\omega t},
 \end{align}
with 
\begin{equation}\label{eq:fn}
    \tilde{f}^{(n)}=
    \begin{cases}
      \frac{1}{R^3}, & \text{if}\ n=0 \\
      \frac{i n \omega\rho J_1\left(x^{\left(n\right)}\right)}{\pi R\left(x^{\left(n\right)} J_0\left(x^{\left(n\right)}\right)-2J_1\left(x^{\left(n\right)}\right)\right)}, & \text{if}\ n\neq0
    \end{cases}
\end{equation}
where $x^{(n)}=i^{3/2}W_o^{(n)}$, and $W_o^{(n)}=R\sqrt{n\omega\rho/\mu}=\sqrt{4n\omega\tau}$ is the Womersley number of the $n^{th}$ mode, a well-defined parameter that characterizes pulsatile flows by relating the driving frequency to viscous effects. In Eq.~\ref{eq:fn}, $J_0$ and $J_1$ are the zeroth and first order Bessel functions of the first kind.


Eqs.~\ref{eq:PQ1} and \ref{eq:PQ2} allow for exact solutions for the axial pressure and current in a vessel as long as their boundary values (either pressure or current) are known. The solutions for $Q$ and $P$ for each vessel can be expressed in terms of $\omega$, $L$ and the values of $\lambda$, $\tau$, and $\alpha$ for that vessel at each time-step of adaptation (see Sec.~A1 of the appendix). The driving signal for radius evolution $\langle \sigma Q\rangle$ can thus be calculated for each vessel in the network, and we obtain for the $e^{th}$ vessel,
\begin{equation}
    \langle\sigma_e Q_e\rangle=\sum_{n=-\infty}^{n=\infty}Re[\tilde{f}_e^{(n)}]\langle Q_e^2\rangle^{(n)},
\end{equation}
where $\langle Q_e^2\rangle^{(n)}$ is the contribution of the $n^{th}$ frequency mode to the mean-squared current $\langle Q_e^2\rangle$ (see Sec.~A2 of the appendix for full functional form of $\langle\sigma_e Q_e\rangle$). Thus the driving signal for the pulsatile-flow rule (Eq.~\ref{eq:AE_P}) differs from that of the steady-flow rule (Eq.~\ref{eq:AE_S}), with the contribution of each frequency mode weighted by the real part of $\tilde{f}$ in the former, which captures the phase-difference between the flow rate and the wall-shear stress on the time-scale of adaptation.
\section{Results}\label{secResults}
\begin{figure*}
\begin{center}
\includegraphics[scale=0.6]{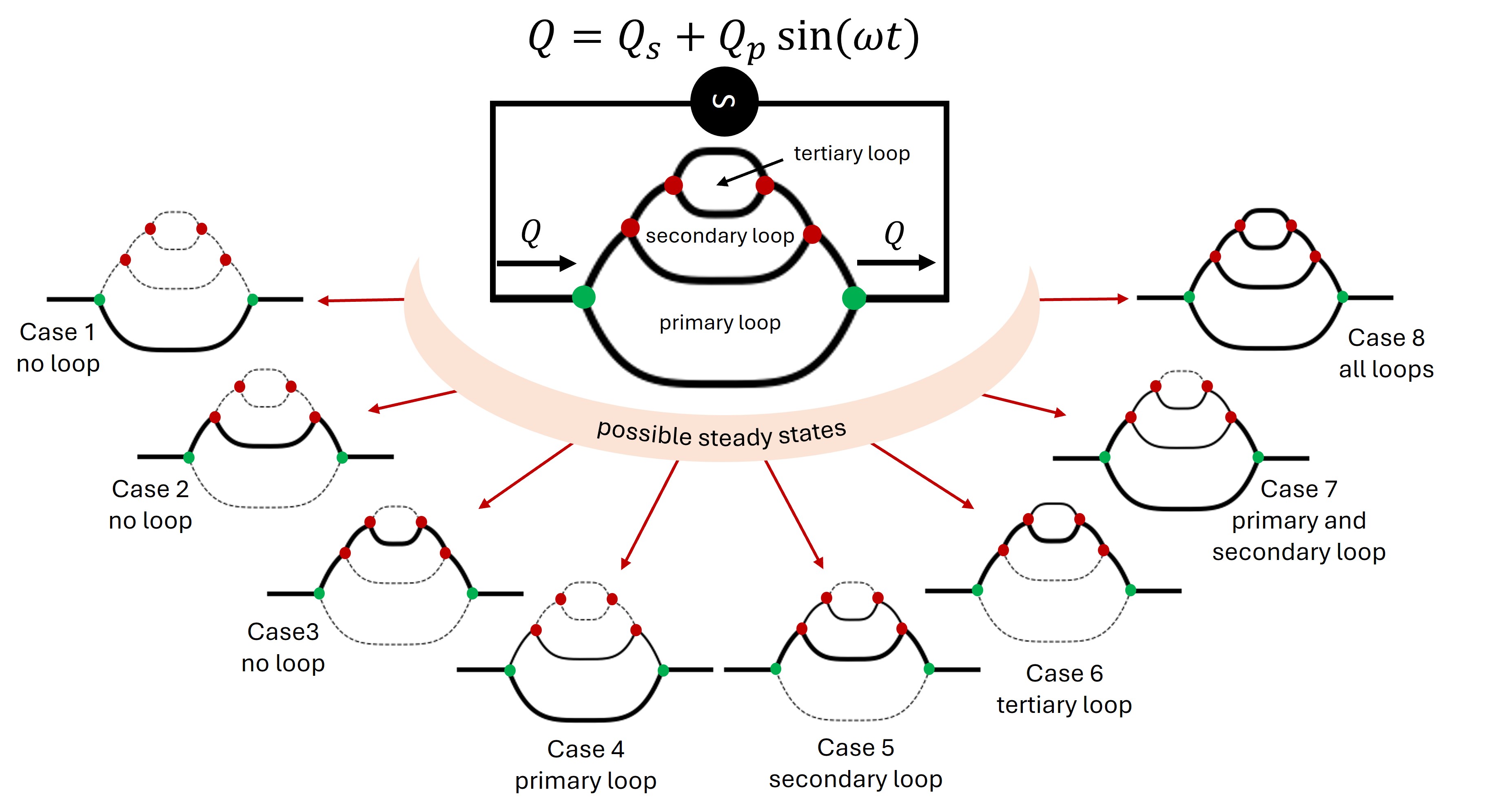}
\end{center}
\caption {\small {Eight vessel toy network with a current $Q$ from the pulsatile source entering and leaving the network through two externally driven nodes (green dots), and  with three levels of looping. Arrows point to possible steady states that may be obtained through long-term adaptation of the network, with dashed lines illustrating vessel shunting, i.e. the vessel radius going to zero.}}
    \label{fig:cartoon}
\end{figure*}
We study adaptation in networks of compliant vessels with different levels of loops, and with a pulsatile source current of the form $Q=Q_s+Q_p \sin{(\omega t)}$ entering and leaving the toy network through two externally driven nodes (Fig.~\ref{fig:cartoon}). Such a choice allows us to mimic the closed circuit comprised by the mammalian ciculatory system driven by the periodically beating heart. For the purposes of this study, we always assume finite pulsatility of the source current (i.e $Q_p\neq 0$) even in the limit of small driving frequency. While we set $Q_s=Q_P=1$ for most of this paper, we will also investigate the effect of varying $Q_p$ while holding $Q_s$ constant (see Sec.~A5 of the appendix). For completeness, we will show that our results are also generalizable to a network architecture with current sinks instead of a closed circuit (see Sec.~A6 of the appendix).

The toy network illustrated in Fig.~\ref{fig:cartoon} is one example with three levels of loops, characterized as primary, secondary or tertiary based on their proximity to the driven nodes (green) of the network. The possibility of vessel shunting and loop destabilization allows for many different long-time structures as depicted through arrows in Fig.~\ref{fig:cartoon}. However, only one of these possible steady states is realized for a given set of initial conditions and for a particular driving frequency $\omega$ and metabolic cost parameter $\gamma$. We are interested in determining the conditions under which loops can be stabilized in such networks. 

A scalar order parameter that characterizes the extent to which the steady-state structure is looped, is the network entropy $S$ \cite{ang2005} given by
\begin{equation}\label{eq:NetEntS}
    S=\sum_{j=0}^{N} P_j S_j, 
\end{equation}
where $P_j=Q_j/Q_s$ is the ratio of the total outflow ($Q_j$) from the $j^{th}$ node, to the net steady-flow source current $Q_s$. $S_j$ is the Shannon entropy of the $j^{th}$ node given by
\begin{equation}\label{eq:Sj}
    S=\sum_{k\in \mathcal{N}} p_{jk} \ln{p_{jk}}. 
\end{equation}
In Eq.~\ref{eq:Sj}, $\mathcal{N}$ is the set of all nodes connected to node $j$ by a single edge, and $p_{jk}=Q_{jk}/Q_j$ is the fraction of the total outflow $Q_j$ that the outflow from node $j$ to node $k$ ($Q_{jk}$) constitutes. Qualitatively, the network entropy $S$ quantifies the path uncertainty in the toy networks under steady-flow conditions: $S=0$ corresponds to there being a single path between the two externally driven nodes (cases $1-3$ in Fig.~\ref{fig:cartoon}), $S=1$ corresponds to every path between the two driven nodes being equally probable (case $8$ in Fig.~\ref{fig:cartoon}), and $0<S<1$ corresponds to some paths between the driven nodes being more probable than others (cases $4-7$ in Fig.~\ref{fig:cartoon}).

In what follows, we will consider three different network architectures: a two-vessel network with only a primary loop, a five-vessel network with a primary and a secondary loop, and an eight-vessel network identical to Fig.~\ref{fig:cartoon} with a primary, secondary and a tertiary loop. For the simplest two-vessel network, we will compare the results of the steady-flow rule in Eq.~\ref{eq:AE_S} and the pulsatile-flow rule in Eq.~\ref{eq:AE_P} for different values of $\gamma$ and $\omega$, and show that they yield qualitatively similar structure phase-diagrams in the $\gamma-\omega\tau_0$ phase-space. For the five and eight vessel networks, we will only report the long-term structures stabilized by the pulsatile-flow rule, and analyze the phenotypic variation of the steady state with the driving frequency and differences in the path-lengths between the externally driven nodes.

\subsection{Two Vessel Network}\label{subsecTwoVessel}
\begin{figure*}
\begin{center}
\includegraphics[scale=0.6]{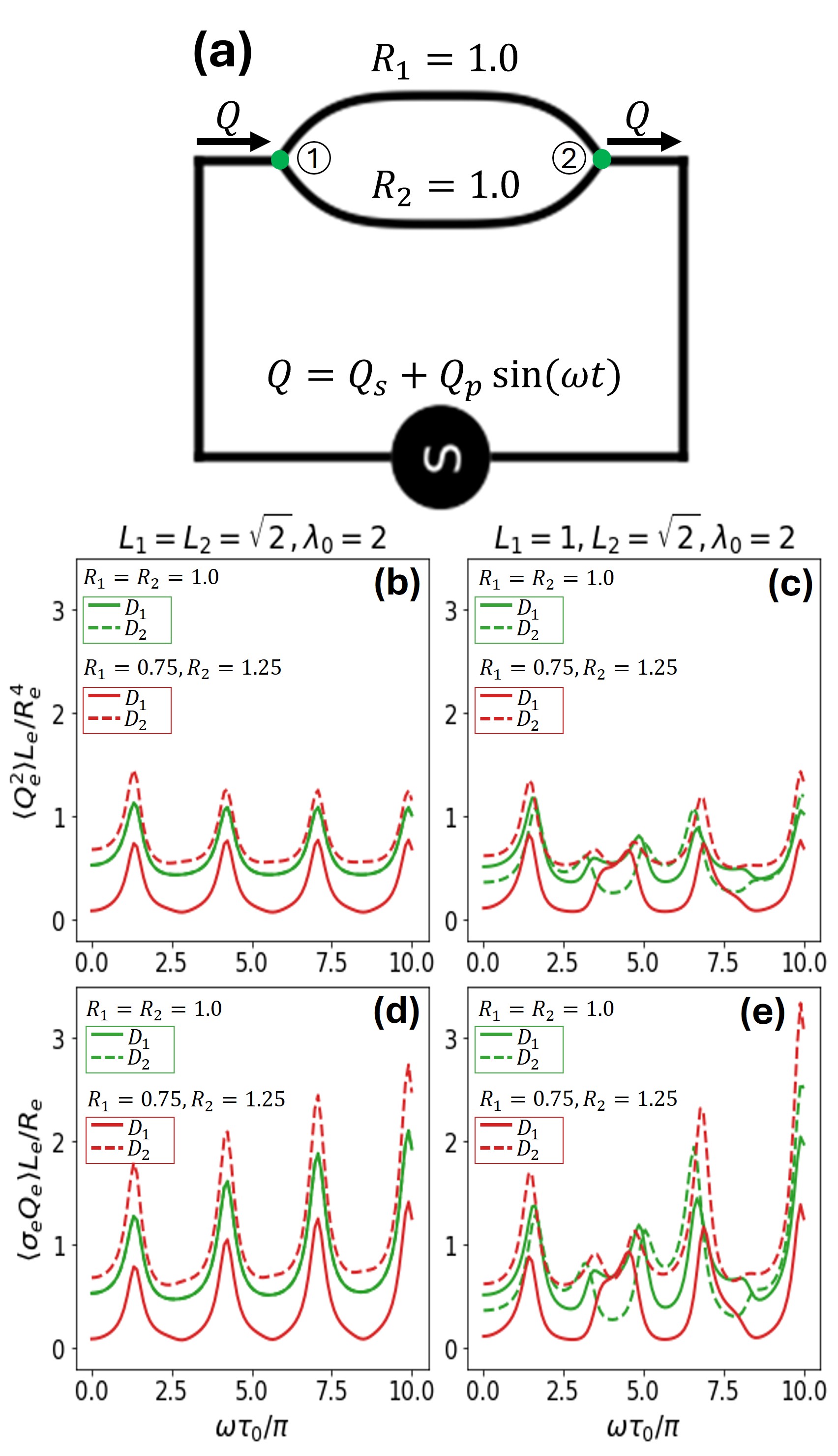}
\end{center}
\caption {(a) Two vessel toy network. (b-e) The energy dissipated per unit time in the $e^{th}$ vessel as a function of the re-scaled frequency $\omega\tau_0$, with $D_e=\langle Q_e^2\rangle L_e/R_e^4$ in the steady-flow case in (b,c) and $D_e=\langle \sigma_e Q_e\rangle L_e/R_e$ in the pulsatile-flow case (d,e). Panels (b) and (d) correspond to vessels of equal lengths $L_1=L_2=1.0$, and panels (c) and (e) correspond to a case of unequal vessel lengths with $L_1=1.0, L_2=\sqrt{2}$. In (b-e), the solid curves correspond to vessel $1$ and the dashed curves to vessel $2$, whereas the two colors differentiate between two distinct sets of radii pairs: green for $R_1=1.0$, $R_2=1.0$ and red for $R_1=0.75$, $R_2=1.25$. In both the steady-flow and pulsatile-flow cases, the dissipation is amplified in each vessel for $\omega\tau_0>0$, with maximum amplification occurring at certain resonant frequencies. The two vessels have identical resonant frequencies only when their lengths are equal, as in (b) and (d). The pulsatile-flow case generally shows higher dissipation amplification for all $\omega\tau_0$ than the steady-flow case, and this trend becomes progressively more pronounced as $\omega\tau_0$ is increased. $R_0=\tau_0=Q_s=Q_p=1$, $\lambda_0=2$.}
    \label{fig:2diss}
\end{figure*}

\begin{figure*}
\begin{center}
\includegraphics[scale=0.75]{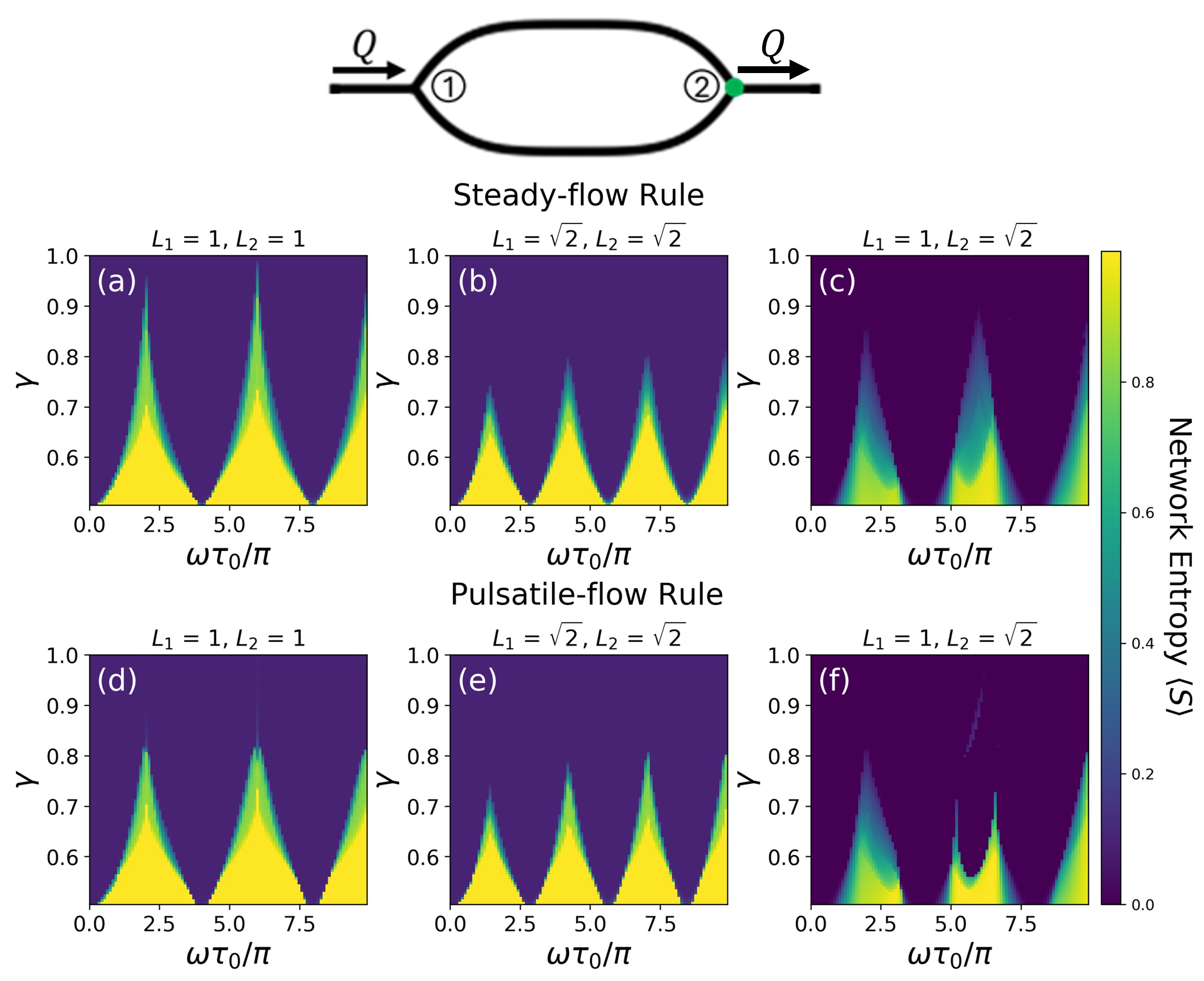}
\end{center}
\caption {Phase diagrams in the $\gamma-\omega\tau_0$ phase-space for the average network entropy $\langle S\rangle$ over $144$ initial conditions, where $\langle S\rangle=1$ corresponds to a symmetric loop ($R_1= R_2$), $0<\langle S\rangle<1$ to an asymmetric loop ($R_1\neq R_2$ and $R_1,R_2>0$) and $\langle S\rangle=0$ to shunting or loop destabilization ($R_1=0$ or $R_2=0$) on average. (a-c) correspond to the steady-flow adaptation rule, with (a,b) considering cases of equal vessel lengths while (c) considers unequal vessel lengths. Similarly (d-f) correspond to the pulsatile-flow rule, with equal vessel lengths considered in (d,e), but not in (f). In all cases, loops are stabilized in broad regions of the phase space with $\gamma>1/2$, with $\langle S\rangle>0$. For vessels of equal lengths (a,b,d,e), the critical value $\gamma_c(\omega\tau_0)$ above which loops become unstable, shows periodic modulations as a function of the re-scaled frequency, reaching maximum values at certain resonant frequencies. An overall increase in the vessel length in comparison to the damping length-scale $\lambda_0$ results in shorter and more frequent resonant peaks, as seen by comparing (a) to (b) and (d) to (e). For unequal vessel lengths, $\gamma_c(\omega\tau_0)$ is not strictly periodic, and loop stabilization can only occur in regions where the resonances of each vessel, which are distinct, have significant overlap. While the steady and pulsatile flow rules yield nearly identical phase diagrams in the equal vessel length cases (a,b) and (d,e), for unequal vessel lengths the two rules generate somewhat different steady-state structures especially in regions of the phase-space where symmetric loops cannot be supported (e.g around $\omega\tau_0=6\pi$). This is because of the heightened asymmetry in dissipation amplification of each vessel under the pulsatile-flow rule, which favors the growth of a thick vessel at the expense of the other, thinner vessel, leading to shunting instead of the stabilization of an asymmetric loop. The equal vessel length cases (a,b,d,e) can always stabilize a symmetric loop when the starting radii of the two vessels are exactly equal, unlike the case of unequal vessel lengths (c,f). This leads to the minimum value of $\langle S\rangle$ to be higher in (a,b,d,e) than in (c,f).  $R_0=\tau_0=Q_s=Q_p=a=b=1$ and $\lambda_0=2$.}
    \label{fig:2phase}
\end{figure*}

Fig.~\ref{fig:2diss}(a) illustrates the two-vessel network with a single loop between the two driven nodes. In all cases the flow is pulsatile. However, the relevant cost function plotted can either follow the steady-flow relevant rule of Eq.~\ref{eq:AE_S}, or the pulsatile-flow relevant rule of Eq.~\ref{eq:AE_P}. Figs.~\ref{fig:2diss}(b,c) show the energy $D_e$ dissipated per unit time in the $e^{th}$ vessel of the two-vessel network for the steady-flow rule case, while Figs.~\ref{fig:2diss}(d,e) show the same for the pulsatile-flow rule case. Figs.~\ref{fig:2diss} (b,d) correspond to vessels of equal lengths $L_1=L_2$, whereas Figs.~\ref{fig:2diss} (c,e) correspond to the case where one vessel is shorter than the other. In both the steady-flow and the pulsatile-flow case, we look at the energy dissipation in each vessel (differentiated by solid and dashed curves) as a function of the re-scaled frequency $\omega\tau_0$, for two different sets of radii $R_1$ and $R_2$ (green or orange) for the two vessels. In each case, we find that the energy dissipated is amplified for $\omega\tau_0 \neq 0$, i.e. when the non-dynamical approximation is relaxed. The amplification of dissipation is maximized at certain resonant frequencies, which are identical in each vessel only when their lengths are equal. 

In comparing Figs.~\ref{fig:2diss}(b,c) with Figs.~\ref{fig:2diss}(d,e) we find that the pulsatile-flow case shows higher dissipation amplification for both vessels and for all $\omega\tau_0$, in comparison to the steady-flow case. Additionally, the amplification of the dissipated energy becomes progressively more pronounced in the pulsatile-flow case as the re-scaled frequency $\omega\tau_0$ increases, unlike in the steady-flow case where all resonant peaks have more or less the same height. Moreover, while the dissipation of the thicker vessel is always larger, the asymmetry in the amplification of dissipation in the two vessels is much larger in the pulsatile-flow case than in the steady-flow case. In other words, the ratio $D_2/D_1$, when vessel $2$ is thicker than vessel $1$, is larger for pulsatile-flow than for steady flow, especially close to resonances.

Fig.~\ref{fig:2phase} shows the phase-diagram of the average network entropy $\langle S\rangle$ in the $\gamma-\omega\tau_0$ phase-space, for the long-time structures generated by the steady-flow adaptation rule (top row) as well as those generated by the pulsatile-flow adaptation rule (bottom row). As mentioned at the start of this section, the network entropy, averaged over many initial sets of radii for the two vessels,  characterizes the ``loopiness" of the steady state network structure, such that $\langle S\rangle=1$ corresponds to a symmetric loop ($R_1=R_2$), $0<\langle S\rangle<1$ to an asymmetric loop ($R_1 \neq R_2$ but $R_1,R_2>0$) and $\langle S\rangle=0$ to loop destabilization or shunting ($R_1=0$ or $R_2=0$), on average. 

When short-term pulsatile dynamics is neglected and flow relaxation in individual vessels is assumed to be instantaneous (i.e. in the non-dynamical limit), Eq.~\ref{eq:AE_S} and Eq.~\ref{eq:AE_P} can both be analytically shown to lead to the destabilization of looped structures for all $\gamma>1/2$ \cite{hu2013, chatterjee2024}. For $\gamma>1/2$ the solution space of these adaptation rules goes from being convex to being non-convex. In other words, they become `winner takes all' rules,  which triggers an instability leading to a negative feedback loop whereby a marginally thicker vessel (higher dissipation) can grow continuously at the expense of a thinner vessel whose dissipation has fallen below a certain threshold. This ultimately leads to the shunting of the thinner vessel, that is, the destabilization of the loop. The only exception to this is in the case where the two vessels have exactly equal starting radii, because they continue to have the same thickness throughout adaptation, preventing the ``winner takes all" instability from being triggered \cite{chatterjee2024}. However, even though two vessels with identical starting radii can support a symmetric loop at steady state for $\gamma>1/2$, this steady state is unstable and can be rapidly destabilized by small perturbations.

The loss of convexity of these rules for $\gamma>1/2$ should intuitively be understood as it being energetically favorable for the cost function\textemdash that balances dissipation and metabolic cost to construct or maintain the vessel\textemdash to route all flow through one big vessel rather than maintaining two thinner vessels.is However, all physiologically relevant metabolic or material costs are characterized by $\gamma$ values greater than $1/2$, suggesting that there must be an additional mechanism to stabilize loops in this range, given their ubiquity in biological networks.

In Fig.~\ref{fig:2phase}, both the steady-flow and the pulsatile-flow adaptation rules show loop stabilization for $\gamma>1/2$ unlike in the non-dynamical approximation ($\omega\tau_0=0$). This can be attributed to the weakening of the `winner takes all' effect in Eqs.~\ref{eq:AE_S} and \ref{eq:AE_P} for $\gamma>1/2$ when short-term pulsatile dynamics of compliant vessels is allowed to determine the long-term remodeling signal. This short-term pulsatility leads to the amplification of the energy dissipated in each vessel, ensuring that when one vessel is marginally thicker than the other, the dissipation of the thinner vessel is large enough to avoid a negative feedback loop leading to an instability and eventual shunting. Thus, it is really the ability to respond to the pulsatile flow in a compliant manner at short time-scales that leads to robust loop stabilization for $\gamma>1/2$ \cite{chatterjee2024}.

The critical value $\gamma_c (\omega\tau_0)$ above which the loop is de-stabilized shows periodic modulations with the re-scaled frequency $\omega\tau_0$ for vessels of equal length, in both the steady and the pulsatile flow cases. At resonant frequencies, for which the amplification of the energy dissipated in each vessel is maximal, loop stabilization is possible for the largest range of $\gamma$ values, i.e. $\gamma_c(\omega\tau_0)$ is maximized. An overall increase in the lengths of equally long vessels, as reflected in going from (a) to (b) or from (d) to (e) in Fig.~\ref{fig:2phase}, results in shorter and more frequent resonant peaks. This is because as the length of a vessel increases in comparison to  the damping length-scale $\lambda_0$, the effect of short-term pulsatile dynamics on its long-term remodeling signal decreases. 

It is to be noted that a symmetric loop can always be stabilized in the case of vessels of equal lengths when both vessels start with identical initial radii. This is true both with the steady-flow and the pulsatile-flow rule, and owing to this special symmetry, the minimum network entropy in the equal vessel length cases (a,b,d,e) is close to $0.1$, because a tenth of the initial radii considered for the average are exactly equal. This is in contrast to the case of unequal vessel lengths (c,f) where the symmetric loop is not always a stable steady state, for which the minimum values for $\langle S\rangle$ are of the order $10^{-16}$.

In comparing Figs.~\ref{fig:2phase}(a-c) and Figs.~\ref{fig:2phase}(d-e) we find that the phase-diagrams corresponding to the steady-flow and the pulsatile-flow adaptation rules only show minor quantitative differences in $\langle S\rangle$ for vessels of equal lengths, while for vessels with unequal lengths these differences are more pronounced. This can be explained by the larger asymmetry in the dissipation amplification of the two vessels with the pulsatile-flow rule, as illustrated in Fig.~\ref{fig:2diss}(d,e), especially close to resonances. This asymmetric dissipation amplification works in favor of the ``winner takes all effect", destabilizing loops where one would have existed if the network was evolved with the steady-flow rule, e.g. at $\omega\tau_0=6\pi$ in Fig.~\ref{fig:2phase}(c,f). 

It must be noted however, that despite the differences between the steady states generated by the steady-flow and the pulsatile-flow adaptation rules in the case of vessels that have different lengths, the broad qualitative features of the network entropy phase-diagrams, including the number and position of resonances, are the same between the two rules. Moreover, we compared them to two other rules not derived from an energy optimization functional, one with a driving signal given by $\langle \sigma_e^2 \rangle^{\gamma}/R_e^3$, and another with $\langle \sigma_e \rangle^{2\gamma}/R_e^3$. We chose these two rules to compare Eq.~\ref{eq:AE_S} and Eq.~\ref{eq:AE_P}, because they contain the simplest driving signals that can be constructed from just the wall shear stress, which we know can be directly sensed in the vessels locally. While the first yields closely related and qualitatively similar long-term structures to Eq.~\ref{eq:AE_S} and Eq.~\ref{eq:AE_P}, the latter where the pulsatile contributions average to zero over one cycle of periodic driving, shows starkly different results (see Sec.~A4 of the appendix). These observations lead us to conclude that the mechanism of loop stabilization is less dependent on the exact rule that governs long-term remodeling, and more on the ability of the rule to capture the short-term pulsatile dynamics of vessels in response to periodic driving. Additionally, they justify our use of the pulsatile rule (Eq.~\ref{eq:AE_P}) for network remodeling, even if its driving signal (involving the flow rate $Q$) cannot be exactly measured in individual vessels. This is because we can expect the actually measured adaptation signal in individual vessels to take the network at least approximately down the gradient of the pulsatile energy optimization functional Eq. ~\ref{eq:EF_P}. We will restrict our study to the pulsatile-flow rule in what follows.

While we hold $Q_s=Q_p=1$ constant for most results reported here, we investigated the effect of varying the pulsatile amplitude $Q_p$ in relation to the steady flow amplitude $Q_s$ for the same two vessel network architecture. Increasing $Q_p$ increases the pulsatile contribution to the driving signal, which further amplifies the net dissipation in each vessel as compared to the results shown in Fig.~\ref{fig:2diss} with $Q_p=1$. Thus higher $Q_p$ values lead to taller and broader resonant peaks in the phase space of $\gamma-\omega\tau_0$ (see Sec.~A5 of the appendix). We also considered a case where the looped network and the pulsatile source do not form a closed circuit as in Fig.~\ref{fig:2diss}(a), but which instead has a sink node from which the steady part of the source current can flow out (see Sec.~A6 of the appendix). We find robust loop stabilization for $\gamma>1/2$ in this case as well, pointing to the generality of our results concerning the importance of short-term pulsatility, irrespective of network architectures and boundary conditions.
\subsection{Five Vessel Network}\label{subsecFiveVessel}
\begin{figure*}
\begin{center}
\includegraphics[scale=0.6]{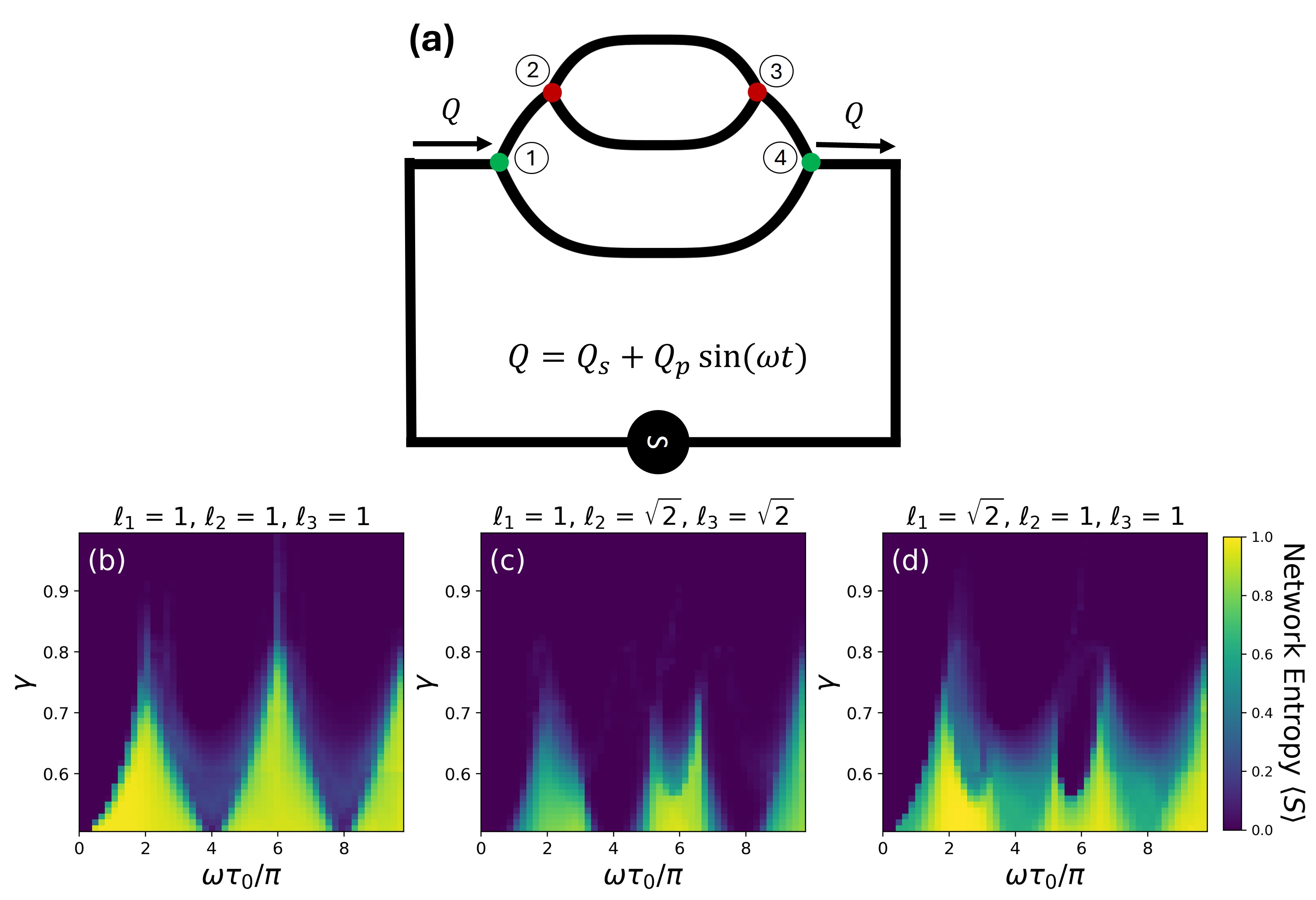}
\end{center}
\caption {(a) Five vessel toy network. (b-d) Phase-diagrams in the $\gamma-\omega\tau_0$ phase-space for the average network entropy $\langle S\rangle$ for $250$ initial conditions. The primary path-length $\ell_1$ corresponds to the length of the single vessel connecting nodes $1$ and $4$, whereas the secondary path-lengths $\ell_2$ and $\ell_3$ correspond to the total length of the two paths that connect node $1$ to $2$ to $3$ to $4$. All three cases show robust loop stabilization for $\gamma>1/2$, although the resonant frequencies at which loops are stabilized for the largest range of $\gamma$ values, are periodically spaced only in the case where all three path-lengths are equal, as in (b). For unequal path-lengths, loop stabilization is stronger when the primary path is longer than the secondary path as in (d). $R_0=\tau_0=Q_s=Q_p=a=b=1$ and $\lambda_0=2$.}
    \label{fig:5phase}
\end{figure*}

\begin{figure*}
\begin{center}
\includegraphics[scale=0.8]{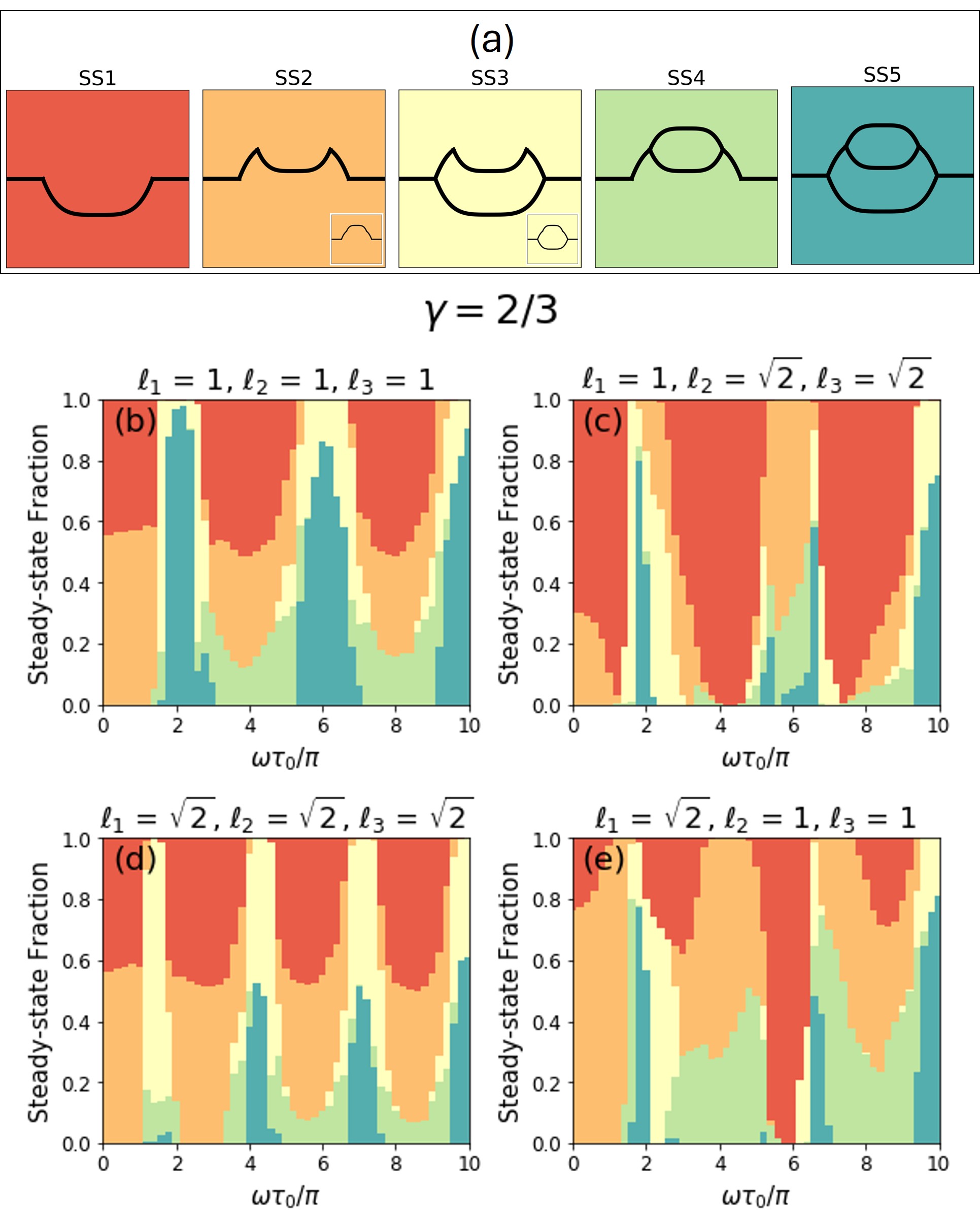}
\end{center}
\caption {(a) All possible steady state ($SS$) architectures allowed for the five vessel network. (b-e) Fraction of each possible structure in $SS1-SS5$ obtained at steady-state for $250$ different initial conditions, for $\gamma=2/3$. The color coding of the steady state fraction follows that of panel (a), e.g. red for $SS1$, orange for $SS2$ etc. When the three path-lengths between the driven nodes are equal, as in (b,d), resonant frequencies are periodically spaced, unlike in the unequal path-length cases in (c,e). In all cases however, resonant frequencies stabilize the fully looped structure ($SS5$) or the primary loop ($SS3$), whereas anti-resonant frequencies lead to loopless structures ($SS1,SS2$), or stabilize the secondary loop ($SS4$). Comparing (c) and (e) we find that when the primary path-length ($\ell_1$) is longer than the secondary path-lengths ($\ell_2,\ell_3$), the probability that the secondary loop ($SS4$) is stabilized close to anti-resonant frequencies, is larger, and the probability of obtaining loopless structures ($SS1,SS2$) is smaller. $R_0=\tau_0=Q_s=Q_p=a=b=1$ and $\lambda_0=2$.}
    \label{fig:5SSfrac}
\end{figure*}
We now consider the five vessel network illustrated in Fig.~\ref{fig:5phase}(a), which has a primary and a secondary loop, and three possible paths between the two driven nodes (green). Denoting each path-length by $\ell_i$, the path directly connecting the two driven nodes ($1\to4$, primary path) has length $\ell_1$ and the two paths involving the vessels of the secondary loop ($1\to2\to3\to4$, secondary paths) have total lengths $\ell_2$ and $\ell_3$. Note that $\ell_2$ and $\ell_3$ specify the total length of the two paths between nodes $1$ and $4$ that involve the secondary loop, and the lengths of the vessels connecting nodes $1$ to $2$, $2$ to $3$ and $3$ to $4$ along these paths are not separately specified. Fig.~\ref{fig:5phase}(b) shows the phase diagram of the average network entropy $\langle S\rangle$ in the $\gamma-\omega\tau_0$ phase space for a few different values of the three path-lengths $\ell_i$. The fully looped structure is stabilized for $\langle S\rangle=1$, only one path between the two driven nodes survive when $\langle S\rangle\approx 0$, and a combination of paths between the driven nodes are possible when $0<\langle S\rangle<1$.

In agreement with results for the two-vessel network, we find that when the path-lengths are equal, the critical value $\gamma_c(\omega\tau_0)$ oscillates as a function of $\omega\tau_0$, and is moreover maximized at resonant frequencies that occur with a well-defined periodicity (Fig.~\ref{fig:5phase}(b)). Robust loop stabilization for $\gamma>1/2$ also occurs for unequal path-lengths, and is stronger when the primary path $\ell_1$ is longer than the secondary paths $\ell_2$ and $\ell_3$, as can be seen from comparing Fig.~\ref{fig:5phase}(c) to Fig.~\ref{fig:5phase}(d). 

Interestingly, the phase-diagrams in Figs.~\ref{fig:5phase}(b,c) are very similar to those in Figs.~\ref{fig:2phase}(d) and (f) in the positioning, width and height of resonances, suggesting that the addition of an extra (secondary) path does not appreciably change the long-time structure. However, because the two available paths between the driven nodes are equivalent in the two-vessel case, the steady state would be symmetric under an interchange of the lengths of the two vessels in Fig.~\ref{fig:2phase}(f). In other words, $L_1=1, L_2=\sqrt{2}$ and $L_1=\sqrt{2}, L_2=1$ would yield identical phase-diagrams for $\langle S\rangle$. The fact that this is not true for the five-vessel network as seen in Figs.~\ref{fig:5phase}(c,d) confirms that the primary and the secondary paths are not equivalent. We highlight this observation to clarify that the similarity in the phase diagrams in Figs.~\ref{fig:2phase}(d,f) and Figs.~\ref{fig:5phase}(b,c) does not imply identical loop-stabilization irrespective of the complexity of the network architecture.

Fig.~\ref{fig:5SSfrac}(a) shows every possible long-time structure (SS$1$-SS$5$) for the five-vessel network, sequentially colored so that going from red to blue we go from the least looped to the maximally looped structure. Figs.~\ref{fig:5SSfrac}(b-e) illustrate the fraction of each structure obtained at steady state for $250$ sets of initial radii, as a function of $\omega\tau_0$ and for $\gamma=2/3$. Figs.~\ref{fig:5SSfrac}(b,d) confirm that resonant (or anti-resonant) frequencies occur with a well-defined periodicity when the path-lengths between the driven nodes are equal. In these two cases, resonant frequencies stabilize either the full-looped structure (SS$5$, blue) or the primary loop (SS$3$, yellow) at steady-state, whereas at anti-resonant frequencies we either get completely loopless structures (SS$1$, red and SS$2$, orange) or only the secondary loop (SS$4$, green). Moreover, an overall increase in path-lengths leads to more closely spaced resonances, and weaker net loop stabilization. This matches our observation with the two-vessel network, that the effect of pulsatility and hence the possibility of loop stabilization is weaker in longer vessels.

Figs.~\ref{fig:5SSfrac}(c,e) confirm our observations from Figs.~\ref{fig:5phase}(c,d) that when the path-lengths are unequal, resonances do not occur with any well-defined periodicity. However, the fully looped structure or the higher level (primary) loop is once again stabilized at these resonant frequencies, whereas away from resonances we get loopless structures or lower level (secondary) loops. Moreover, comparing Figs.~\ref{fig:5SSfrac}(c) and (e) we find that the probability for the secondary loop (SS$4$, green) to be stabilized away from resonances is larger when the primary path $\ell_1$ is longer than the secondary paths $\ell_2, \ell_3$. This further explains the larger looped $0<\langle S\rangle<1$ area in Fig.~\ref{fig:5phase}(d) as compared to Fig.~\ref{fig:5phase}(c) as arising through the shunting of the longer primary path (higher resistance) in favor of the shorter secondary paths, and thereby preserving the secondary loop more often on average.

\subsection{Eight Vessel Network}\label{subsecEightVessel}
\begin{figure*}
\begin{center}
\includegraphics[scale=0.65]{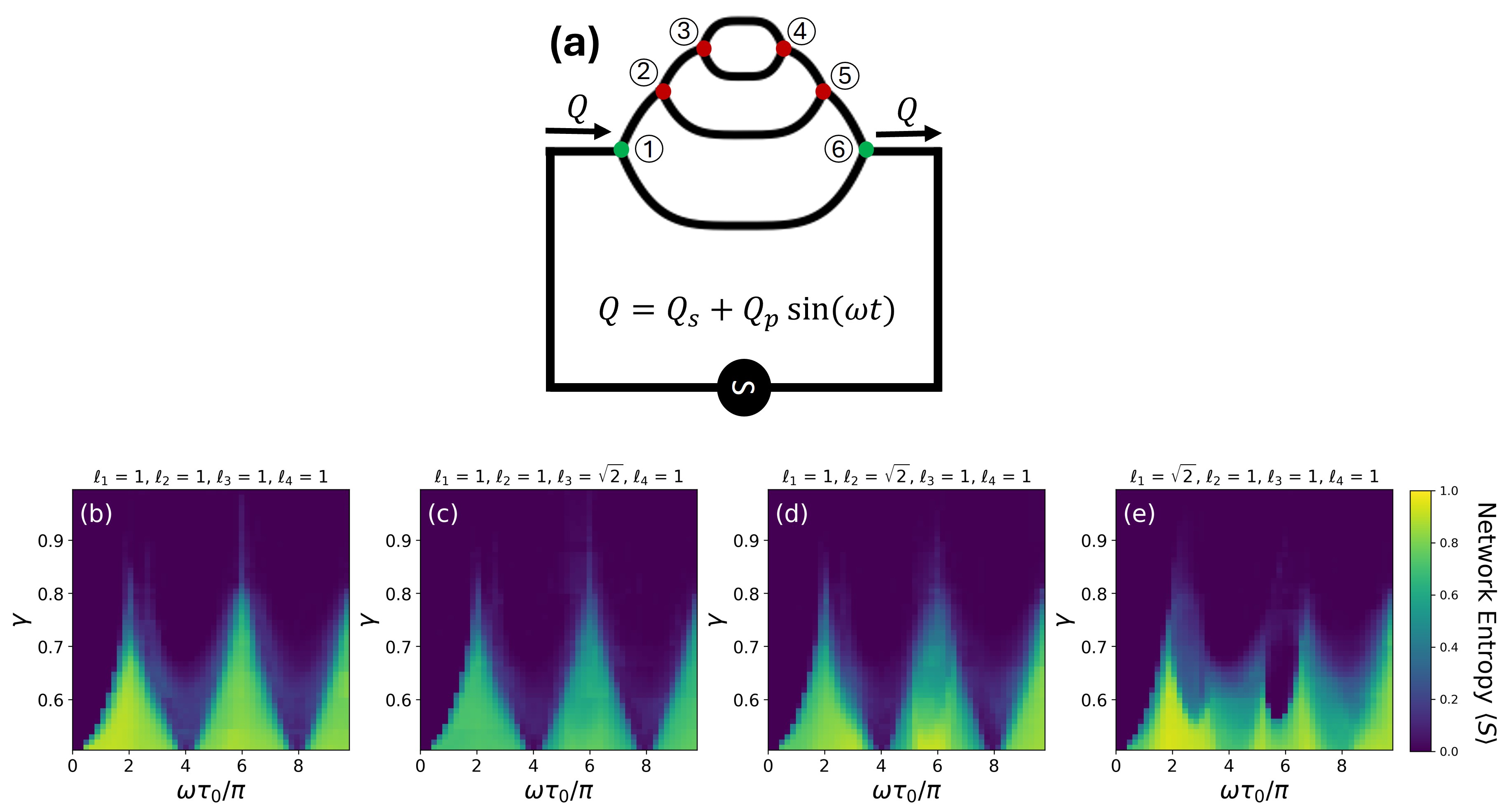}
\end{center}
\caption {(a) Eight vessel toy network. (b-d) Phase-diagrams in the $\gamma-\omega\tau_0$ phase-space for the average network entropy $\langle S\rangle$ for $250$ initial conditions. The primary path-length $\ell_1$ corresponds to the length of the single vessel connecting nodes $1$ and $6$, the secondary path-length $\ell_2$ corresponds to the total length of the path that connects $1\to 2\to 5\to 6$, and the tertiary path-lengths $\ell_3$ and $\ell_4$ correspond to the total lengths of the two paths connecting $1\to 2\to 3\to 4\to 5\to 6$ . All four cases show robust loop stabilization for $\gamma>1/2$, and resonant frequencies at which loops are stabilized for the largest range of $\gamma$ values, are periodically spaced except in (e) where the primary path-length is the longest. Going from (c) to (e) we observe progressively stronger stabilization of looped structures as the longest path goes from being one of the tertiary paths to the secondary path to finally the primary path. $R_0=\tau_0=Q_s=Q_p=a=b=1$ and $\lambda_0=2$.}
    \label{fig:8phase}
\end{figure*}

\begin{figure*}
\begin{center}
\includegraphics[scale=0.8]{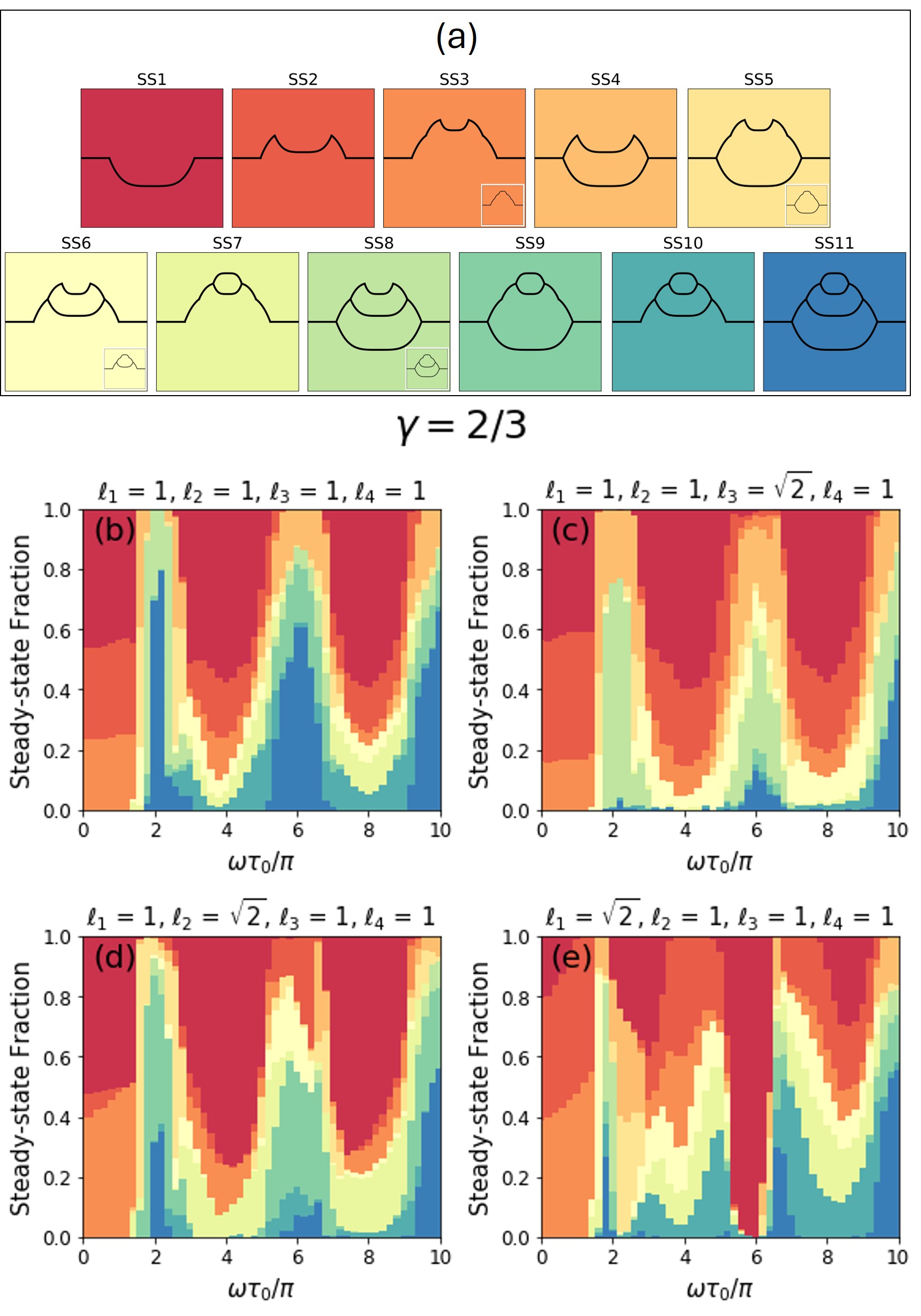}
\end{center}
\caption {(a) All possible steady states for the eight vessel network. (b-e) Fraction of each possible structure in $SS1-SS11$ obtained at steady-state for $250$ different initial conditions, for $\gamma=2/3$. The color coding of the steady state fraction follows that of panel (a), e.g. red for $SS1$, dark orange for $SS2$ etc. Resonant frequencies are more or less periodic except when the primary path $\ell_1$ is the longest, as in (e). In all cases, the fully looped structure, and the primary loop by itself or in combination with the secondary and tertiary loops are stabilized close to resonances. Away from resonances we see shunting, lower level loops, or a combination of lower level loops. In (c) where one of the tertiary paths is longer than the other paths, the lower level loop stabilized away from resonances is the tertiary loop, whereas in (d), where the secondary path is the longest, it is the tertiary loop that is stabilized away from resonances. For (e), where the primary path is the longest, the primary loop is harder to stabilize near resonances than in the other three cases considered, whereas away from resonances both the secondary and tertiary loops by themselves, but also their combination can be stabilized with ease. $R_0=\tau_0=Q_s=Q_p=a=b=1$ and $\lambda_0=2$.}
    \label{fig:8SSfrac}
\end{figure*}

Next we look at the eight vessel network illustrated in Fig.~\ref{fig:8phase}(a) with a primary, a secondary and a tertiary loop, and four possible paths leading from one source node to the other ($1$ and $6$). The primary path connects the source nodes directly ($1\to 6$) and has length $\ell_1$, the secondary path connects the source nodes through one branch of the secondary loop ($1\to 2\to 5 \to 6$) and has total length $\ell_2$ (sum of lengths of all vessels participating in this path), and the two tertiary paths ($1\to 2\to 3\to 4\to 5\to 6$) involve the two branches of the tertiary loop and have total lengths $\ell_3$ and $\ell_4$. The long-time structures of this eight-vessel network, evolved through the pulsatile-flow adaptation rule (Eq.~\ref{eq:AE_P}), show qualitatively similar features to those of the two-vessel and five-vessel networks. Here we will restrict our analysis to a few descriptive cases of path-length asymmetry to further highlight our conclusions from the other two network architectures. 

Figs.~\ref{fig:8phase}(b-e) show the $\gamma-\omega\tau_0$ phase diagram of the average network entropy $\langle S\rangle$, and compare the steady-state structures when all path-lengths are equal (b) to cases when one of the tertiary paths (c), or the secondary path (d) or the primary path (e) are longer than the other paths between the two driven nodes. In all cases, we find loop stabilization for $\gamma>1/2$, and except in Fig.~\ref{fig:8phase}(e), the critical value $\gamma_c(\omega\tau_0)$ above which looped structures are destabilized shows more or less periodic modulations with $\omega\tau_0$, peaking at resonant frequencies. For unequal path-lengths we observe a progressive increase in the area of the phase-space where looped structures are stable as we go from (c) to (e), or in other words, as the longest path goes from being one of the tertiary paths, to the secondary path and finally the primary path. 

Fig.~\ref{fig:8SSfrac}(a) shows every possible long-time structure SS$1$-SS$11$ for the eight-vessel network. Figs.~\ref{fig:8SSfrac}(b-e) illustrate the fraction of each of these structures obtained at steady state for $250$ initial conditions color coded according to Fig.~\ref{fig:8SSfrac}(a), as a function of $\omega\tau_0$ and for $\gamma=2/3$, for the path-length combinations illustrated in Fig.~\ref{fig:8phase}(b-e). For the equal path-length case (b) resonances occur periodically, and at resonant frequencies the most encountered steady states in descending order of probability are the fully looped structure (SS$11$), the primary loop (SS$4$) or a combination of the primary and lower level loops (SS$9$ and SS$8$). At anti-resonant frequencies, we get mostly loopless structures (SS$1$, SS$2$ and SS$3$), a small fraction of steady states with only the secondary (SS$6$) or the tertiary (SS$7$) loops preserved, and an even smaller fraction of steady-states with a combination of the secondary and tertiary loops stabilized (SS$10$). The fact that at these anti-resonant frequencies, loopless structures are more likely to have the primary path intact over the secondary or tertiary paths suggests that it is energetically more favorable to maintain a single long vessel than multiple vessels connecting the two driven nodes.

Figs.~\ref{fig:8SSfrac}(c-e) show qualitatively similar features, with fully looped structures, higher-level loops, or combinations involving higher level loops prioritized for stabilization near resonances, whereas loopless structures, lower-level loops, or combinations involving lower-level loops favored away from resonances. What is different is that in these cases of unequal path-lengths, resonant frequencies also stabilize lower level loops or combinations like SS$10$ and SS$6$, albeit with a lower probability. In (c), where one of the tertiary paths is longer than the other paths, the lower-level loop stabilized away from resonances is the secondary loop ($SS6$), whereas the tertiary loop can only be stabilized in combination with the primary ($SS9$) or secondary loop ($SS10$) near resonances, and almost never on its own. Moreover, loopless structures that do not involve the longer tertiary path are preferred at anti-resonant frequencies. In contrast, for (d) where the secondary path is the longest, it is the tertiary loop ($SS7$) that is preferred for stabilization away from resonances, and the secondary loop (on its own or in combination with the tertiary and primary loops, like in $SS6$, $SS8$ and $SS10$) can only be stabilized close to resonant frequencies. The loopless structure involving the secondary path has very low probability of occurrence near anti-resonant frequencies.

In contrast to the cases with longer secondary and tertiary paths (Figs.~\ref{fig:8SSfrac}(c,d)), resonances lack periodicity when the primary path is the longest as in Fig.~\ref{fig:8SSfrac}(e). In this case we observe shunting ($SS1$-$SS3$), stand-alone secondary ($SS6$) and tertiary loops ($SS7$) or a combination of the two ($SS10$) away from resonances, while the primary loop (alone as in $SS5$ or in combination as in $SS8$ and $SS9$) is only stabilized near resonances. Comparing (e) with (c) and (d) in Fig.~\ref{fig:8SSfrac} we find that the observed increase in the looped areas of the $\langle S\rangle$ phase-diagram away from resonances when the primary path is longer than the secondary or tertiary paths (Fig.~\ref{fig:8phase}(d)), arises from the increased dominance of the steady state with both the secondary and tertiary loops preserved (SS$10$).
\section{Discussion}\label{sec:discussion}
In summary, we have introduced a new adaptation rule, showcased  in Eq.~\ref{eq:AE_P},  for networks of compliant vessels. This rule is more realistic for remodeling in the presence of pulsatile driving, and emerges directly from dissipation minimization in dynamic conditions. We have compared the steady-state structures generated by this remodeling rule with those produced by a widely used quasi-steady flow rule for animal and plant vascular development (Eq.~\ref{eq:AE_S}) for a minimal two-vessel network (Fig.~\ref{fig:cartoon}). We showed that when short-term pulsatile dynamics in individual compliant vessels is accounted for in the long-term remodeling signal, both these rules allow for robust loop stabilization. This loop stabilization is a result of resonances that amplify dissipation in each vessel at certain frequencies, and takes place for a broad range of physiologically relevant metabolic costs where existing theories of adaptation would predict loopless structures. This is equally true of networks that do not form a closed circuit with the pulsatile source, as is the case for all results reported in the main text, but also for cases with current sinks (see Sec.~A6 of the appendix). 

The qualitative features of the structure phase-diagram for different values of the driving frequency and the metabolic cost appear to be more or less independent of the exact remodeling rule, whether they are derivable from an energy optimization functional or not (see Sec.~A4 of the appendix). This suggests that loop stabilization for physiologically relevant metabolic costs depends primarily on the ability of compliant vessels to respond dynamically to the pulsatile flow at short time-scales. This dynamical response is affected by geometric considerations, i.e. length of paths that will be stabilized or shunted, coupling tightly the network geometry to the stabilized topology. As the phenomenon of loop stabilization is generally robust to the details of the use-it-or-lose-it remodeling rule, we expect it to be of importance in a large class of living networks that operate and remodel under pulsatile conditions.

Moreover, hierarchical networks with several levels of loops, when evolved through the pulsatile-flow remodeling rule show considerable variability in their long-time structures. Consolidating the results from all three network architectures considered in this study, we find the following general observations. Firstly, higher-level loops proximal to the source, by themselves or in combination with lower-level loops, are prioritized for stabilization at or near resonant frequencies. In contrast, away from resonances it seems to be energetically more favorable to shunt one or more paths between the driven nodes, leading to loopless structures or preserving only the lower-level loops distal from the source (or combinations of them). 

Secondly, path-length asymmetries play an important role in determining the network entropy, i.e. the overall `loopiness' of the steady-state structure. For the same radius, longer path-lengths are associated with higher net dissipation. When paths that are distal to the source (secondary or tertiary paths) are the longest in the network, there is a decreased probability of stabilizing lower-level loops away from resonances, because it is energetically favorable to shunt these longer paths. In contrast, when the path most directly connected to the driven nodes (primary path) is longer than the others, lower-level loops can be preserved away from resonances, making the steady-state more 'loopy' (larger values of $\langle S\rangle$) in general in the phase-space of  $\gamma-\omega\tau_0$.

Thus, depending on the driving frequency, and the lengths of the various available paths between the source nodes, very different architectures are possible, ranging from more branched structures with lower level loops away from resonances, and reticulated structures with higher level loops close to resonant frequencies. We conclude that the mechanism of loop stabilization through resonances arising from short-term pulsatility, can offer an explanation for the wide variety of structural phenotypes observed in biology, even if our minimal networks do not capture the complexity of actual biological networks and flows. 

 For the sake of simplicity, we have assumed that the flow velocity in individual compliant vessels has a parabolic profile, which allows for exact solutions of the axial current and pressure as a function of time. In reality however, the flow in compliant vessels driven by pulsatile sources has Womersley profiles that deviate from the parabolic approximation more strongly as the driving frequency increases. While we relegate the construction and study of more realistic flow-dynamics models and network architectures to future work, the results presented here suggest that loop-stabilization through short-term pulsatile dynamics is a robust and general feature of periodically driven adapting networks of compliant vessels, which would carry over to improved models.

 In a broader context, this work highlighted how an adaptive network, remodeling under local rules, can adopt a variety of loopy architectures due to the coupling of the vessel elasticity to the flow pulsatility and the emergence of resonances. This complex landscape of network architectures results from a competition between geometry (length of paths) and topology (path redundancy) and is influenced by the driving frequency of the power source. Given the robustness of the effect to the details of the remodeling rule, we believe that the underlying physics (pulsatile driving and resonance amplification) could be of interest in the context of learning in physical networks, in addition to its importance for vascular physiology.

\begin{acknowledgments}
The authors acknowledge support from the Simons Foundation through Award 568888, the University  of Pennsylvania  Materials  Research  Science and  Engineering  Center  (MRSEC)  through Grant No. DMR-1720530 and DMR-2309043, the John Templeton Foundation by award number 62846 and the HFSP Award 977405.
\end{acknowledgments}

\section*{Data Availability}
The data that support the findings of this article are openly available~\cite{chatterjee2025data}.

\bibliographystyle{apsrev4-2}
\bibliography{hierarchical_loop_stabilization}

\begin{thebibliography}{61}%
\makeatletter
\providecommand \@ifxundefined [1]{%
 \@ifx{#1\undefined}
}%
\providecommand \@ifnum [1]{%
 \ifnum #1\expandafter \@firstoftwo
 \else \expandafter \@secondoftwo
 \fi
}%
\providecommand \@ifx [1]{%
 \ifx #1\expandafter \@firstoftwo
 \else \expandafter \@secondoftwo
 \fi
}%
\providecommand \natexlab [1]{#1}%
\providecommand \enquote  [1]{``#1''}%
\providecommand \bibnamefont  [1]{#1}%
\providecommand \bibfnamefont [1]{#1}%
\providecommand \citenamefont [1]{#1}%
\providecommand \href@noop [0]{\@secondoftwo}%
\providecommand \href [0]{\begingroup \@sanitize@url \@href}%
\providecommand \@href[1]{\@@startlink{#1}\@@href}%
\providecommand \@@href[1]{\endgroup#1\@@endlink}%
\providecommand \@sanitize@url [0]{\catcode `\\12\catcode `\$12\catcode
  `\&12\catcode `\#12\catcode `\^12\catcode `\_12\catcode `\%12\relax}%
\providecommand \@@startlink[1]{}%
\providecommand \@@endlink[0]{}%
\providecommand \url  [0]{\begingroup\@sanitize@url \@url }%
\providecommand \@url [1]{\endgroup\@href {#1}{\urlprefix }}%
\providecommand \urlprefix  [0]{URL }%
\providecommand \Eprint [0]{\href }%
\providecommand \doibase [0]{https://doi.org/}%
\providecommand \selectlanguage [0]{\@gobble}%
\providecommand \bibinfo  [0]{\@secondoftwo}%
\providecommand \bibfield  [0]{\@secondoftwo}%
\providecommand \translation [1]{[#1]}%
\providecommand \BibitemOpen [0]{}%
\providecommand \bibitemStop [0]{}%
\providecommand \bibitemNoStop [0]{.\EOS\space}%
\providecommand \EOS [0]{\spacefactor3000\relax}%
\providecommand \BibitemShut  [1]{\csname bibitem#1\endcsname}%
\let\auto@bib@innerbib\@empty
\bibitem [{\citenamefont {Nelson}\ and\ \citenamefont
  {Dengler}(1997)}]{Nelson1997}%
  \BibitemOpen
  \bibfield  {author} {\bibinfo {author} {\bibfnamefont {T.}~\bibnamefont
  {Nelson}}\ and\ \bibinfo {author} {\bibfnamefont {N.}~\bibnamefont
  {Dengler}},\ }\href {https://doi.org/10.1105/tpc.9.7.1121} {\bibfield
  {journal} {\bibinfo  {journal} {The Plant Cell}\ }\textbf {\bibinfo {volume}
  {9}},\ \bibinfo {pages} {1121} (\bibinfo {year} {1997})}\BibitemShut
  {NoStop}%
\bibitem [{\citenamefont {Laguna}\ \emph {et~al.}(2008)\citenamefont {Laguna},
  \citenamefont {Bohn},\ and\ \citenamefont {Jagla}}]{laguna2008}%
  \BibitemOpen
  \bibfield  {author} {\bibinfo {author} {\bibfnamefont {M.~F.}\ \bibnamefont
  {Laguna}}, \bibinfo {author} {\bibfnamefont {S.}~\bibnamefont {Bohn}},\ and\
  \bibinfo {author} {\bibfnamefont {E.~A.}\ \bibnamefont {Jagla}},\ }\href@noop
  {} {\bibfield  {journal} {\bibinfo  {journal} {PLoS computational biology}\
  }\textbf {\bibinfo {volume} {4}},\ \bibinfo {pages} {e1000055} (\bibinfo
  {year} {2008})}\BibitemShut {NoStop}%
\bibitem [{\citenamefont {Ronellenfitsch}\ and\ \citenamefont
  {Katifori}(2019)}]{ronellenfitsch2019}%
  \BibitemOpen
  \bibfield  {author} {\bibinfo {author} {\bibfnamefont {H.}~\bibnamefont
  {Ronellenfitsch}}\ and\ \bibinfo {author} {\bibfnamefont {E.}~\bibnamefont
  {Katifori}},\ }\href@noop {} {\bibfield  {journal} {\bibinfo  {journal}
  {Physical review letters}\ }\textbf {\bibinfo {volume} {123}},\ \bibinfo
  {pages} {248101} (\bibinfo {year} {2019})}\BibitemShut {NoStop}%
\bibitem [{\citenamefont {Katifori}\ \emph {et~al.}(2010)\citenamefont
  {Katifori}, \citenamefont {Sz{\"o}ll{\H{o}}si},\ and\ \citenamefont
  {Magnasco}}]{katifori2010}%
  \BibitemOpen
  \bibfield  {author} {\bibinfo {author} {\bibfnamefont {E.}~\bibnamefont
  {Katifori}}, \bibinfo {author} {\bibfnamefont {G.~J.}\ \bibnamefont
  {Sz{\"o}ll{\H{o}}si}},\ and\ \bibinfo {author} {\bibfnamefont {M.~O.}\
  \bibnamefont {Magnasco}},\ }\href@noop {} {\bibfield  {journal} {\bibinfo
  {journal} {Physical review letters}\ }\textbf {\bibinfo {volume} {104}},\
  \bibinfo {pages} {048704} (\bibinfo {year} {2010})}\BibitemShut {NoStop}%
\bibitem [{\citenamefont {Corson}(2010)}]{corson2010}%
  \BibitemOpen
  \bibfield  {author} {\bibinfo {author} {\bibfnamefont {F.}~\bibnamefont
  {Corson}},\ }\href@noop {} {\bibfield  {journal} {\bibinfo  {journal}
  {Physical Review Letters}\ }\textbf {\bibinfo {volume} {104}},\ \bibinfo
  {pages} {048703} (\bibinfo {year} {2010})}\BibitemShut {NoStop}%
\bibitem [{\citenamefont {Kaiser}\ \emph {et~al.}(2020)\citenamefont {Kaiser},
  \citenamefont {Ronellenfitsch},\ and\ \citenamefont {Witthaut}}]{kaiser2020}%
  \BibitemOpen
  \bibfield  {author} {\bibinfo {author} {\bibfnamefont {F.}~\bibnamefont
  {Kaiser}}, \bibinfo {author} {\bibfnamefont {H.}~\bibnamefont
  {Ronellenfitsch}},\ and\ \bibinfo {author} {\bibfnamefont {D.}~\bibnamefont
  {Witthaut}},\ }\href@noop {} {\bibfield  {journal} {\bibinfo  {journal}
  {Nature communications}\ }\textbf {\bibinfo {volume} {11}},\ \bibinfo {pages}
  {5796} (\bibinfo {year} {2020})}\BibitemShut {NoStop}%
\bibitem [{\citenamefont {Pries}\ \emph {et~al.}(1998)\citenamefont {Pries},
  \citenamefont {Secomb},\ and\ \citenamefont {Gaehtgens}}]{pries1998}%
  \BibitemOpen
  \bibfield  {author} {\bibinfo {author} {\bibfnamefont {A.}~\bibnamefont
  {Pries}}, \bibinfo {author} {\bibfnamefont {T.}~\bibnamefont {Secomb}},\ and\
  \bibinfo {author} {\bibfnamefont {P.}~\bibnamefont {Gaehtgens}},\ }\href@noop
  {} {\bibfield  {journal} {\bibinfo  {journal} {American Journal of
  Physiology-Heart and Circulatory Physiology}\ }\textbf {\bibinfo {volume}
  {275}},\ \bibinfo {pages} {H349} (\bibinfo {year} {1998})}\BibitemShut
  {NoStop}%
\bibitem [{\citenamefont {Pries}\ and\ \citenamefont
  {Secomb}(2008)}]{pries2008}%
  \BibitemOpen
  \bibfield  {author} {\bibinfo {author} {\bibfnamefont {A.~R.}\ \bibnamefont
  {Pries}}\ and\ \bibinfo {author} {\bibfnamefont {T.~W.}\ \bibnamefont
  {Secomb}},\ }\href@noop {} {\bibfield  {journal} {\bibinfo  {journal}
  {Microcirculation}\ }\textbf {\bibinfo {volume} {15}},\ \bibinfo {pages}
  {753} (\bibinfo {year} {2008})}\BibitemShut {NoStop}%
\bibitem [{\citenamefont {Hu}\ and\ \citenamefont {Cai}(2013)}]{hu2013}%
  \BibitemOpen
  \bibfield  {author} {\bibinfo {author} {\bibfnamefont {D.}~\bibnamefont
  {Hu}}\ and\ \bibinfo {author} {\bibfnamefont {D.}~\bibnamefont {Cai}},\
  }\href@noop {} {\bibfield  {journal} {\bibinfo  {journal} {Physical review
  letters}\ }\textbf {\bibinfo {volume} {111}},\ \bibinfo {pages} {138701}
  (\bibinfo {year} {2013})}\BibitemShut {NoStop}%
\bibitem [{\citenamefont {Ronellenfitsch}\ and\ \citenamefont
  {Katifori}(2016)}]{ronellenfitsch2016}%
  \BibitemOpen
  \bibfield  {author} {\bibinfo {author} {\bibfnamefont {H.}~\bibnamefont
  {Ronellenfitsch}}\ and\ \bibinfo {author} {\bibfnamefont {E.}~\bibnamefont
  {Katifori}},\ }\href@noop {} {\bibfield  {journal} {\bibinfo  {journal}
  {Physical review letters}\ }\textbf {\bibinfo {volume} {117}},\ \bibinfo
  {pages} {138301} (\bibinfo {year} {2016})}\BibitemShut {NoStop}%
\bibitem [{\citenamefont {Chang}\ and\ \citenamefont
  {Roper}(2019)}]{chang2019}%
  \BibitemOpen
  \bibfield  {author} {\bibinfo {author} {\bibfnamefont {S.-S.}\ \bibnamefont
  {Chang}}\ and\ \bibinfo {author} {\bibfnamefont {M.}~\bibnamefont {Roper}},\
  }\href@noop {} {\bibfield  {journal} {\bibinfo  {journal} {Journal of
  theoretical biology}\ }\textbf {\bibinfo {volume} {462}},\ \bibinfo {pages}
  {48} (\bibinfo {year} {2019})}\BibitemShut {NoStop}%
\bibitem [{\citenamefont {Gounaris}\ \emph {et~al.}(2024)\citenamefont
  {Gounaris}, \citenamefont {Jovchevska}, \citenamefont {Garcia},\ and\
  \citenamefont {Katifori}}]{gounaris2024central}%
  \BibitemOpen
  \bibfield  {author} {\bibinfo {author} {\bibfnamefont {G.}~\bibnamefont
  {Gounaris}}, \bibinfo {author} {\bibfnamefont {M.}~\bibnamefont
  {Jovchevska}}, \bibinfo {author} {\bibfnamefont {M.~R.}\ \bibnamefont
  {Garcia}},\ and\ \bibinfo {author} {\bibfnamefont {E.}~\bibnamefont
  {Katifori}},\ }\href {https://arxiv.org/abs/2111.04657} {\bibinfo {title}
  {The central role of metabolism in vascular morphogenesis}} (\bibinfo {year}
  {2024}),\ \Eprint {https://arxiv.org/abs/2111.04657} {arXiv:2111.04657
  [q-bio.TO]} \BibitemShut {NoStop}%
\bibitem [{\citenamefont {Kramer}\ and\ \citenamefont
  {Modes}(2023)}]{kramer2023}%
  \BibitemOpen
  \bibfield  {author} {\bibinfo {author} {\bibfnamefont {F.}~\bibnamefont
  {Kramer}}\ and\ \bibinfo {author} {\bibfnamefont {C.~D.}\ \bibnamefont
  {Modes}},\ }\href@noop {} {\bibfield  {journal} {\bibinfo  {journal}
  {Physical Review Research}\ }\textbf {\bibinfo {volume} {5}},\ \bibinfo
  {pages} {023106} (\bibinfo {year} {2023})}\BibitemShut {NoStop}%
\bibitem [{\citenamefont {Marbach}\ \emph {et~al.}(2023)\citenamefont
  {Marbach}, \citenamefont {Ziethen},\ and\ \citenamefont
  {Alim}}]{marbach2023}%
  \BibitemOpen
  \bibfield  {author} {\bibinfo {author} {\bibfnamefont {S.}~\bibnamefont
  {Marbach}}, \bibinfo {author} {\bibfnamefont {N.}~\bibnamefont {Ziethen}},\
  and\ \bibinfo {author} {\bibfnamefont {K.}~\bibnamefont {Alim}},\ }\href@noop
  {} {\bibfield  {journal} {\bibinfo  {journal} {New Journal of Physics}\
  }\textbf {\bibinfo {volume} {25}},\ \bibinfo {pages} {123052} (\bibinfo
  {year} {2023})}\BibitemShut {NoStop}%
\bibitem [{\citenamefont {Travasso}\ \emph {et~al.}(2025)\citenamefont
  {Travasso}, \citenamefont {Penick}, \citenamefont {Dunn},\ and\ \citenamefont
  {Poir{\'e}}}]{travasso2025predicting}%
  \BibitemOpen
  \bibfield  {author} {\bibinfo {author} {\bibfnamefont {R.~D.}\ \bibnamefont
  {Travasso}}, \bibinfo {author} {\bibfnamefont {C.~A.}\ \bibnamefont
  {Penick}}, \bibinfo {author} {\bibfnamefont {R.~R.}\ \bibnamefont {Dunn}},\
  and\ \bibinfo {author} {\bibfnamefont {E.~C.}\ \bibnamefont {Poir{\'e}}},\
  }\href@noop {} {\bibfield  {journal} {\bibinfo  {journal} {Scientific
  Reports}\ }\textbf {\bibinfo {volume} {15}},\ \bibinfo {pages} {7017}
  (\bibinfo {year} {2025})}\BibitemShut {NoStop}%
\bibitem [{\citenamefont {Harazny}\ \emph {et~al.}(2014)\citenamefont
  {Harazny}, \citenamefont {Ott}, \citenamefont {Raff}, \citenamefont
  {Welzenbach}, \citenamefont {Kwella}, \citenamefont {Michelson},\ and\
  \citenamefont {Schmieder}}]{harazny2014}%
  \BibitemOpen
  \bibfield  {author} {\bibinfo {author} {\bibfnamefont {J.~M.}\ \bibnamefont
  {Harazny}}, \bibinfo {author} {\bibfnamefont {C.}~\bibnamefont {Ott}},
  \bibinfo {author} {\bibfnamefont {U.}~\bibnamefont {Raff}}, \bibinfo {author}
  {\bibfnamefont {J.}~\bibnamefont {Welzenbach}}, \bibinfo {author}
  {\bibfnamefont {N.}~\bibnamefont {Kwella}}, \bibinfo {author} {\bibfnamefont
  {G.}~\bibnamefont {Michelson}},\ and\ \bibinfo {author} {\bibfnamefont
  {R.~E.}\ \bibnamefont {Schmieder}},\ }\href@noop {} {\bibfield  {journal}
  {\bibinfo  {journal} {Journal of hypertension}\ }\textbf {\bibinfo {volume}
  {32}},\ \bibinfo {pages} {2246} (\bibinfo {year} {2014})}\BibitemShut
  {NoStop}%
\bibitem [{\citenamefont {Fancher}\ and\ \citenamefont
  {Katifori}(2022)}]{fancher2022}%
  \BibitemOpen
  \bibfield  {author} {\bibinfo {author} {\bibfnamefont {S.}~\bibnamefont
  {Fancher}}\ and\ \bibinfo {author} {\bibfnamefont {E.}~\bibnamefont
  {Katifori}},\ }\href@noop {} {\bibfield  {journal} {\bibinfo  {journal}
  {Physical Review Fluids}\ }\textbf {\bibinfo {volume} {7}},\ \bibinfo {pages}
  {013101} (\bibinfo {year} {2022})}\BibitemShut {NoStop}%
\bibitem [{\citenamefont {Fancher}\ and\ \citenamefont
  {Katifori}(2023)}]{fancher2023tradeoff}%
  \BibitemOpen
  \bibfield  {author} {\bibinfo {author} {\bibfnamefont {S.}~\bibnamefont
  {Fancher}}\ and\ \bibinfo {author} {\bibfnamefont {E.}~\bibnamefont
  {Katifori}},\ }\href {https://arxiv.org/abs/2102.13197} {\bibinfo {title}
  {Tradeoffs between energy efficiency and mechanical response in fluid flow
  networks}} (\bibinfo {year} {2023}),\ \Eprint
  {https://arxiv.org/abs/2102.13197} {arXiv:2102.13197 [physics.bio-ph]}
  \BibitemShut {NoStop}%
\bibitem [{\citenamefont {Cheng}\ \emph {et~al.}(2014)\citenamefont {Cheng},
  \citenamefont {Williamitis},\ and\ \citenamefont {Slaughter}}]{cheng2014}%
  \BibitemOpen
  \bibfield  {author} {\bibinfo {author} {\bibfnamefont {A.}~\bibnamefont
  {Cheng}}, \bibinfo {author} {\bibfnamefont {C.~A.}\ \bibnamefont
  {Williamitis}},\ and\ \bibinfo {author} {\bibfnamefont {M.~S.}\ \bibnamefont
  {Slaughter}},\ }\href@noop {} {\bibfield  {journal} {\bibinfo  {journal}
  {Annals of cardiothoracic surgery}\ }\textbf {\bibinfo {volume} {3}},\
  \bibinfo {pages} {573} (\bibinfo {year} {2014})}\BibitemShut {NoStop}%
\bibitem [{\citenamefont {Bartoli}\ \emph {et~al.}(2010)\citenamefont
  {Bartoli}, \citenamefont {Giridharan}, \citenamefont {Litwak}, \citenamefont
  {Sobieski}, \citenamefont {Prabhu}, \citenamefont {Slaughter},\ and\
  \citenamefont {Koenig}}]{bartoli2010}%
  \BibitemOpen
  \bibfield  {author} {\bibinfo {author} {\bibfnamefont {C.~R.}\ \bibnamefont
  {Bartoli}}, \bibinfo {author} {\bibfnamefont {G.~A.}\ \bibnamefont
  {Giridharan}}, \bibinfo {author} {\bibfnamefont {K.~N.}\ \bibnamefont
  {Litwak}}, \bibinfo {author} {\bibfnamefont {M.}~\bibnamefont {Sobieski}},
  \bibinfo {author} {\bibfnamefont {S.~D.}\ \bibnamefont {Prabhu}}, \bibinfo
  {author} {\bibfnamefont {M.~S.}\ \bibnamefont {Slaughter}},\ and\ \bibinfo
  {author} {\bibfnamefont {S.~C.}\ \bibnamefont {Koenig}},\ }\href@noop {}
  {\bibfield  {journal} {\bibinfo  {journal} {Asaio Journal}\ }\textbf
  {\bibinfo {volume} {56}},\ \bibinfo {pages} {410} (\bibinfo {year}
  {2010})}\BibitemShut {NoStop}%
\bibitem [{\citenamefont {Muhire}\ \emph {et~al.}(2019)\citenamefont {Muhire},
  \citenamefont {Iulita}, \citenamefont {Vallerand}, \citenamefont {Youwakim},
  \citenamefont {Gratuze}, \citenamefont {Petry}, \citenamefont {Planel},
  \citenamefont {Ferland},\ and\ \citenamefont {Girouard}}]{muhire2019}%
  \BibitemOpen
  \bibfield  {author} {\bibinfo {author} {\bibfnamefont {G.}~\bibnamefont
  {Muhire}}, \bibinfo {author} {\bibfnamefont {M.~F.}\ \bibnamefont {Iulita}},
  \bibinfo {author} {\bibfnamefont {D.}~\bibnamefont {Vallerand}}, \bibinfo
  {author} {\bibfnamefont {J.}~\bibnamefont {Youwakim}}, \bibinfo {author}
  {\bibfnamefont {M.}~\bibnamefont {Gratuze}}, \bibinfo {author} {\bibfnamefont
  {F.~R.}\ \bibnamefont {Petry}}, \bibinfo {author} {\bibfnamefont
  {E.}~\bibnamefont {Planel}}, \bibinfo {author} {\bibfnamefont
  {G.}~\bibnamefont {Ferland}},\ and\ \bibinfo {author} {\bibfnamefont
  {H.}~\bibnamefont {Girouard}},\ }\href@noop {} {\bibfield  {journal}
  {\bibinfo  {journal} {Journal of the American Heart Association}\ }\textbf
  {\bibinfo {volume} {8}},\ \bibinfo {pages} {e011630} (\bibinfo {year}
  {2019})}\BibitemShut {NoStop}%
\bibitem [{\citenamefont {Gr{\"a}wer}\ \emph {et~al.}(2015)\citenamefont
  {Gr{\"a}wer}, \citenamefont {Modes}, \citenamefont {Magnasco},\ and\
  \citenamefont {Katifori}}]{grawer2015}%
  \BibitemOpen
  \bibfield  {author} {\bibinfo {author} {\bibfnamefont {J.}~\bibnamefont
  {Gr{\"a}wer}}, \bibinfo {author} {\bibfnamefont {C.~D.}\ \bibnamefont
  {Modes}}, \bibinfo {author} {\bibfnamefont {M.~O.}\ \bibnamefont
  {Magnasco}},\ and\ \bibinfo {author} {\bibfnamefont {E.}~\bibnamefont
  {Katifori}},\ }\href@noop {} {\bibfield  {journal} {\bibinfo  {journal}
  {Physical Review E}\ }\textbf {\bibinfo {volume} {92}},\ \bibinfo {pages}
  {012801} (\bibinfo {year} {2015})}\BibitemShut {NoStop}%
\bibitem [{\citenamefont {Waszkiewicz}\ \emph {et~al.}(2024)\citenamefont
  {Waszkiewicz}, \citenamefont {Shaw}, \citenamefont {Lisicki},\ and\
  \citenamefont {Szymczak}}]{waszkiewicz2024}%
  \BibitemOpen
  \bibfield  {author} {\bibinfo {author} {\bibfnamefont {R.}~\bibnamefont
  {Waszkiewicz}}, \bibinfo {author} {\bibfnamefont {J.~B.}\ \bibnamefont
  {Shaw}}, \bibinfo {author} {\bibfnamefont {M.}~\bibnamefont {Lisicki}},\ and\
  \bibinfo {author} {\bibfnamefont {P.}~\bibnamefont {Szymczak}},\ }\href@noop
  {} {\bibfield  {journal} {\bibinfo  {journal} {Physical Review Letters}\
  }\textbf {\bibinfo {volume} {132}},\ \bibinfo {pages} {137401} (\bibinfo
  {year} {2024})}\BibitemShut {NoStop}%
\bibitem [{\citenamefont {Gyllingberg}\ \emph {et~al.}(2025)\citenamefont
  {Gyllingberg}, \citenamefont {Tian},\ and\ \citenamefont
  {Sumpter}}]{gyllingberg2025minimal}%
  \BibitemOpen
  \bibfield  {author} {\bibinfo {author} {\bibfnamefont {L.}~\bibnamefont
  {Gyllingberg}}, \bibinfo {author} {\bibfnamefont {Y.}~\bibnamefont {Tian}},\
  and\ \bibinfo {author} {\bibfnamefont {D.~J.}\ \bibnamefont {Sumpter}},\
  }\href@noop {} {\bibfield  {journal} {\bibinfo  {journal} {Journal of the
  Royal Society Interface}\ }\textbf {\bibinfo {volume} {22}} (\bibinfo {year}
  {2025})}\BibitemShut {NoStop}%
\bibitem [{\citenamefont {Chatterjee}\ \emph {et~al.}(2024)\citenamefont
  {Chatterjee}, \citenamefont {Fancher},\ and\ \citenamefont
  {Katifori}}]{chatterjee2024}%
  \BibitemOpen
  \bibfield  {author} {\bibinfo {author} {\bibfnamefont {P.}~\bibnamefont
  {Chatterjee}}, \bibinfo {author} {\bibfnamefont {S.}~\bibnamefont
  {Fancher}},\ and\ \bibinfo {author} {\bibfnamefont {E.}~\bibnamefont
  {Katifori}},\ }\href@noop {} {\bibfield  {journal} {\bibinfo  {journal}
  {Physical Review Research}\ }\textbf {\bibinfo {volume} {6}},\ \bibinfo
  {pages} {043015} (\bibinfo {year} {2024})}\BibitemShut {NoStop}%
\bibitem [{\citenamefont {Tero}\ \emph {et~al.}(2010)\citenamefont {Tero},
  \citenamefont {Saigusa}, \citenamefont {Kobayashi},\ and\ \citenamefont
  {Nakagaki}}]{tero2010}%
  \BibitemOpen
  \bibfield  {author} {\bibinfo {author} {\bibfnamefont {A.}~\bibnamefont
  {Tero}}, \bibinfo {author} {\bibfnamefont {T.}~\bibnamefont {Saigusa}},
  \bibinfo {author} {\bibfnamefont {R.}~\bibnamefont {Kobayashi}},\ and\
  \bibinfo {author} {\bibfnamefont {T.}~\bibnamefont {Nakagaki}},\ }\href@noop
  {} {\bibfield  {journal} {\bibinfo  {journal} {Science}\ }\textbf {\bibinfo
  {volume} {327}},\ \bibinfo {pages} {439} (\bibinfo {year}
  {2010})}\BibitemShut {NoStop}%
\bibitem [{\citenamefont {Hacking}\ \emph {et~al.}(1996)\citenamefont
  {Hacking}, \citenamefont {VanBavel},\ and\ \citenamefont
  {Spaan}}]{hacking1996}%
  \BibitemOpen
  \bibfield  {author} {\bibinfo {author} {\bibfnamefont {W.~J.}\ \bibnamefont
  {Hacking}}, \bibinfo {author} {\bibfnamefont {E.}~\bibnamefont {VanBavel}},\
  and\ \bibinfo {author} {\bibfnamefont {J.~A.~E.}\ \bibnamefont {Spaan}},\
  }\href@noop {} {\bibfield  {journal} {\bibinfo  {journal} {Am. J. Physiol.
  270 (1), H364}\ } (\bibinfo {year} {1996})}\BibitemShut {NoStop}%
\bibitem [{\citenamefont {Rolland-Lagan}\ and\ \citenamefont
  {Prusinkiewicz}(2005)}]{rollandLagan2005}%
  \BibitemOpen
  \bibfield  {author} {\bibinfo {author} {\bibfnamefont {A.~G.}\ \bibnamefont
  {Rolland-Lagan}}\ and\ \bibinfo {author} {\bibfnamefont {P.}~\bibnamefont
  {Prusinkiewicz}},\ }\href@noop {} {\bibfield  {journal} {\bibinfo  {journal}
  {Plant J. 44 (5), 854}\ } (\bibinfo {year} {2005})}\BibitemShut {NoStop}%
\bibitem [{\citenamefont {van Berkel}\ \emph {et~al.}(2013)\citenamefont {van
  Berkel}, \citenamefont {de~Boer}, \citenamefont {Scheres},\ and\
  \citenamefont {ten Tusscher}}]{berkel2013}%
  \BibitemOpen
  \bibfield  {author} {\bibinfo {author} {\bibfnamefont {K.}~\bibnamefont {van
  Berkel}}, \bibinfo {author} {\bibfnamefont {R.~J.}\ \bibnamefont {de~Boer}},
  \bibinfo {author} {\bibfnamefont {B.}~\bibnamefont {Scheres}},\ and\ \bibinfo
  {author} {\bibfnamefont {K.}~\bibnamefont {ten Tusscher}},\ }\href@noop {}
  {\bibfield  {journal} {\bibinfo  {journal} {Development 140 (11), 2253}\ }
  (\bibinfo {year} {2013})}\BibitemShut {NoStop}%
\bibitem [{\citenamefont {Murray}(1926)}]{murray1926physiological}%
  \BibitemOpen
  \bibfield  {author} {\bibinfo {author} {\bibfnamefont {C.~D.}\ \bibnamefont
  {Murray}},\ }\href@noop {} {\bibfield  {journal} {\bibinfo  {journal} {The
  Journal of general physiology}\ }\textbf {\bibinfo {volume} {9}},\ \bibinfo
  {pages} {835} (\bibinfo {year} {1926})}\BibitemShut {NoStop}%
\bibitem [{\citenamefont {Sherman}(1981)}]{sherman1981connecting}%
  \BibitemOpen
  \bibfield  {author} {\bibinfo {author} {\bibfnamefont {T.~F.}\ \bibnamefont
  {Sherman}},\ }\href@noop {} {\bibfield  {journal} {\bibinfo  {journal} {The
  Journal of general physiology}\ }\textbf {\bibinfo {volume} {78}},\ \bibinfo
  {pages} {431} (\bibinfo {year} {1981})}\BibitemShut {NoStop}%
\bibitem [{\citenamefont {Kassab}\ and\ \citenamefont
  {Fung}(1995)}]{kassab1995pattern}%
  \BibitemOpen
  \bibfield  {author} {\bibinfo {author} {\bibfnamefont {G.~S.}\ \bibnamefont
  {Kassab}}\ and\ \bibinfo {author} {\bibfnamefont {Y.-C.~B.}\ \bibnamefont
  {Fung}},\ }\href@noop {} {\bibfield  {journal} {\bibinfo  {journal} {Annals
  of biomedical engineering}\ }\textbf {\bibinfo {volume} {23}},\ \bibinfo
  {pages} {13} (\bibinfo {year} {1995})}\BibitemShut {NoStop}%
\bibitem [{\citenamefont {Zamir}\ \emph {et~al.}(1979)\citenamefont {Zamir},
  \citenamefont {Medeiros},\ and\ \citenamefont
  {Cunningham}}]{zamir1979arterial}%
  \BibitemOpen
  \bibfield  {author} {\bibinfo {author} {\bibfnamefont {M.}~\bibnamefont
  {Zamir}}, \bibinfo {author} {\bibfnamefont {J.}~\bibnamefont {Medeiros}},\
  and\ \bibinfo {author} {\bibfnamefont {T.}~\bibnamefont {Cunningham}},\
  }\href@noop {} {\bibfield  {journal} {\bibinfo  {journal} {The Journal of
  general physiology}\ }\textbf {\bibinfo {volume} {74}},\ \bibinfo {pages}
  {537} (\bibinfo {year} {1979})}\BibitemShut {NoStop}%
\bibitem [{\citenamefont {Kassab}(2006)}]{kassab2006scaling}%
  \BibitemOpen
  \bibfield  {author} {\bibinfo {author} {\bibfnamefont {G.~S.}\ \bibnamefont
  {Kassab}},\ }\href@noop {} {\bibfield  {journal} {\bibinfo  {journal}
  {American Journal of Physiology-Heart and Circulatory Physiology}\ }\textbf
  {\bibinfo {volume} {290}},\ \bibinfo {pages} {H894} (\bibinfo {year}
  {2006})}\BibitemShut {NoStop}%
\bibitem [{\citenamefont {Huo}\ and\ \citenamefont
  {Kassab}(2009)}]{huo2009scaling}%
  \BibitemOpen
  \bibfield  {author} {\bibinfo {author} {\bibfnamefont {Y.}~\bibnamefont
  {Huo}}\ and\ \bibinfo {author} {\bibfnamefont {G.~S.}\ \bibnamefont
  {Kassab}},\ }\href@noop {} {\bibfield  {journal} {\bibinfo  {journal}
  {Biophysical journal}\ }\textbf {\bibinfo {volume} {96}},\ \bibinfo {pages}
  {339} (\bibinfo {year} {2009})}\BibitemShut {NoStop}%
\bibitem [{\citenamefont {Taylor}\ \emph {et~al.}(2024)\citenamefont {Taylor},
  \citenamefont {Saxton}, \citenamefont {Halliday}, \citenamefont {Newman},
  \citenamefont {Hose}, \citenamefont {Kassab}, \citenamefont {Gunn},\ and\
  \citenamefont {Morris}}]{taylor2024systematic}%
  \BibitemOpen
  \bibfield  {author} {\bibinfo {author} {\bibfnamefont {D.~J.}\ \bibnamefont
  {Taylor}}, \bibinfo {author} {\bibfnamefont {H.}~\bibnamefont {Saxton}},
  \bibinfo {author} {\bibfnamefont {I.}~\bibnamefont {Halliday}}, \bibinfo
  {author} {\bibfnamefont {T.}~\bibnamefont {Newman}}, \bibinfo {author}
  {\bibfnamefont {D.}~\bibnamefont {Hose}}, \bibinfo {author} {\bibfnamefont
  {G.~S.}\ \bibnamefont {Kassab}}, \bibinfo {author} {\bibfnamefont {J.~P.}\
  \bibnamefont {Gunn}},\ and\ \bibinfo {author} {\bibfnamefont {P.~D.}\
  \bibnamefont {Morris}},\ }\href@noop {} {\bibfield  {journal} {\bibinfo
  {journal} {American Journal of Physiology-Heart and Circulatory Physiology}\
  }\textbf {\bibinfo {volume} {327}},\ \bibinfo {pages} {H182} (\bibinfo {year}
  {2024})}\BibitemShut {NoStop}%
\bibitem [{\citenamefont {Zhou}\ \emph {et~al.}(1999)\citenamefont {Zhou},
  \citenamefont {Kassab},\ and\ \citenamefont {Molloi}}]{zhou1999design}%
  \BibitemOpen
  \bibfield  {author} {\bibinfo {author} {\bibfnamefont {Y.}~\bibnamefont
  {Zhou}}, \bibinfo {author} {\bibfnamefont {G.~S.}\ \bibnamefont {Kassab}},\
  and\ \bibinfo {author} {\bibfnamefont {S.}~\bibnamefont {Molloi}},\
  }\href@noop {} {\bibfield  {journal} {\bibinfo  {journal} {Physics in
  Medicine \& Biology}\ }\textbf {\bibinfo {volume} {44}},\ \bibinfo {pages}
  {2929} (\bibinfo {year} {1999})}\BibitemShut {NoStop}%
\bibitem [{\citenamefont {Kamiya}\ and\ \citenamefont
  {Togawa}(1972)}]{kamiya1972optimal}%
  \BibitemOpen
  \bibfield  {author} {\bibinfo {author} {\bibfnamefont {A.}~\bibnamefont
  {Kamiya}}\ and\ \bibinfo {author} {\bibfnamefont {T.}~\bibnamefont
  {Togawa}},\ }\href@noop {} {\bibfield  {journal} {\bibinfo  {journal} {The
  Bulletin of mathematical biophysics}\ }\textbf {\bibinfo {volume} {34}},\
  \bibinfo {pages} {431} (\bibinfo {year} {1972})}\BibitemShut {NoStop}%
\bibitem [{\citenamefont {Mayrovitz}\ and\ \citenamefont
  {Roy}(1983)}]{mayrovitz1983microvascular}%
  \BibitemOpen
  \bibfield  {author} {\bibinfo {author} {\bibfnamefont {H.~N.}\ \bibnamefont
  {Mayrovitz}}\ and\ \bibinfo {author} {\bibfnamefont {J.}~\bibnamefont
  {Roy}},\ }\href@noop {} {\bibfield  {journal} {\bibinfo  {journal} {American
  Journal of Physiology-Heart and Circulatory Physiology}\ }\textbf {\bibinfo
  {volume} {245}},\ \bibinfo {pages} {H1031} (\bibinfo {year}
  {1983})}\BibitemShut {NoStop}%
\bibitem [{\citenamefont {Taber}\ \emph {et~al.}(2001)\citenamefont {Taber},
  \citenamefont {Ng}, \citenamefont {Quesnel}, \citenamefont {Whatman},\ and\
  \citenamefont {Carmen}}]{taber2001investigating}%
  \BibitemOpen
  \bibfield  {author} {\bibinfo {author} {\bibfnamefont {L.~A.}\ \bibnamefont
  {Taber}}, \bibinfo {author} {\bibfnamefont {S.}~\bibnamefont {Ng}}, \bibinfo
  {author} {\bibfnamefont {A.~M.}\ \bibnamefont {Quesnel}}, \bibinfo {author}
  {\bibfnamefont {J.}~\bibnamefont {Whatman}},\ and\ \bibinfo {author}
  {\bibfnamefont {C.~J.}\ \bibnamefont {Carmen}},\ }\href@noop {} {\bibfield
  {journal} {\bibinfo  {journal} {Journal of biomechanics}\ }\textbf {\bibinfo
  {volume} {34}},\ \bibinfo {pages} {121} (\bibinfo {year} {2001})}\BibitemShut
  {NoStop}%
\bibitem [{\citenamefont {Hughes}(2015)}]{hughes2015optimality}%
  \BibitemOpen
  \bibfield  {author} {\bibinfo {author} {\bibfnamefont {A.~D.}\ \bibnamefont
  {Hughes}},\ }\href@noop {} {\bibfield  {journal} {\bibinfo  {journal} {Artery
  research}\ }\textbf {\bibinfo {volume} {10}},\ \bibinfo {pages} {1} (\bibinfo
  {year} {2015})}\BibitemShut {NoStop}%
\bibitem [{\citenamefont {Silva}\ and\ \citenamefont
  {Reis}(2015)}]{silva2015scaling}%
  \BibitemOpen
  \bibfield  {author} {\bibinfo {author} {\bibfnamefont {C.}~\bibnamefont
  {Silva}}\ and\ \bibinfo {author} {\bibfnamefont {A.~H.}\ \bibnamefont
  {Reis}},\ }\href@noop {} {\bibfield  {journal} {\bibinfo  {journal}
  {International Journal of Thermal Sciences}\ }\textbf {\bibinfo {volume}
  {88}},\ \bibinfo {pages} {77} (\bibinfo {year} {2015})}\BibitemShut {NoStop}%
\bibitem [{\citenamefont {Painter}\ \emph {et~al.}(2006)\citenamefont
  {Painter}, \citenamefont {Ed{\'e}n},\ and\ \citenamefont
  {Bengtsson}}]{painter2006pulsatile}%
  \BibitemOpen
  \bibfield  {author} {\bibinfo {author} {\bibfnamefont {P.~R.}\ \bibnamefont
  {Painter}}, \bibinfo {author} {\bibfnamefont {P.}~\bibnamefont {Ed{\'e}n}},\
  and\ \bibinfo {author} {\bibfnamefont {H.-U.}\ \bibnamefont {Bengtsson}},\
  }\href@noop {} {\bibfield  {journal} {\bibinfo  {journal} {Theoretical
  Biology and Medical Modelling}\ }\textbf {\bibinfo {volume} {3}},\ \bibinfo
  {pages} {1} (\bibinfo {year} {2006})}\BibitemShut {NoStop}%
\bibitem [{\citenamefont {Taylor}(1967)}]{taylor1967elastic}%
  \BibitemOpen
  \bibfield  {author} {\bibinfo {author} {\bibfnamefont {M.~G.}\ \bibnamefont
  {Taylor}},\ }\href@noop {} {\bibfield  {journal} {\bibinfo  {journal}
  {Gastroenterology}\ }\textbf {\bibinfo {volume} {52}},\ \bibinfo {pages}
  {358} (\bibinfo {year} {1967})}\BibitemShut {NoStop}%
\bibitem [{\citenamefont {Shumal}\ \emph {et~al.}(2025)\citenamefont {Shumal},
  \citenamefont {Saghafian}, \citenamefont {Shirani},\ and\ \citenamefont
  {Nili-AhmadAbadi}}]{shumal2025association}%
  \BibitemOpen
  \bibfield  {author} {\bibinfo {author} {\bibfnamefont {M.}~\bibnamefont
  {Shumal}}, \bibinfo {author} {\bibfnamefont {M.}~\bibnamefont {Saghafian}},
  \bibinfo {author} {\bibfnamefont {E.}~\bibnamefont {Shirani}},\ and\ \bibinfo
  {author} {\bibfnamefont {M.}~\bibnamefont {Nili-AhmadAbadi}},\ }\href@noop {}
  {\bibfield  {journal} {\bibinfo  {journal} {Computers in Biology and
  Medicine}\ }\textbf {\bibinfo {volume} {186}},\ \bibinfo {pages} {109741}
  (\bibinfo {year} {2025})}\BibitemShut {NoStop}%
\bibitem [{\citenamefont {Hahn}\ \emph {et~al.}(2008)\citenamefont {Hahn},
  \citenamefont {Gwon}, \citenamefont {Kwon}, \citenamefont {Choi},
  \citenamefont {Choi}, \citenamefont {Lee}, \citenamefont {Hong},
  \citenamefont {Park},\ and\ \citenamefont {Kim}}]{hahn2008comparison}%
  \BibitemOpen
  \bibfield  {author} {\bibinfo {author} {\bibfnamefont {J.-Y.}\ \bibnamefont
  {Hahn}}, \bibinfo {author} {\bibfnamefont {H.-C.}\ \bibnamefont {Gwon}},
  \bibinfo {author} {\bibfnamefont {S.~U.}\ \bibnamefont {Kwon}}, \bibinfo
  {author} {\bibfnamefont {S.-H.}\ \bibnamefont {Choi}}, \bibinfo {author}
  {\bibfnamefont {J.-H.}\ \bibnamefont {Choi}}, \bibinfo {author}
  {\bibfnamefont {S.~H.}\ \bibnamefont {Lee}}, \bibinfo {author} {\bibfnamefont
  {K.-P.}\ \bibnamefont {Hong}}, \bibinfo {author} {\bibfnamefont {J.~E.}\
  \bibnamefont {Park}},\ and\ \bibinfo {author} {\bibfnamefont {D.~K.}\
  \bibnamefont {Kim}},\ }\href@noop {} {\bibfield  {journal} {\bibinfo
  {journal} {Atherosclerosis}\ }\textbf {\bibinfo {volume} {201}},\ \bibinfo
  {pages} {326} (\bibinfo {year} {2008})}\BibitemShut {NoStop}%
\bibitem [{\citenamefont {Schoenenberger}\ \emph {et~al.}(2012)\citenamefont
  {Schoenenberger}, \citenamefont {Urbanek}, \citenamefont {Toggweiler},
  \citenamefont {Seelos}, \citenamefont {Jamshidi}, \citenamefont {Resink},\
  and\ \citenamefont {Erne}}]{schoenenberger2012deviation}%
  \BibitemOpen
  \bibfield  {author} {\bibinfo {author} {\bibfnamefont {A.~W.}\ \bibnamefont
  {Schoenenberger}}, \bibinfo {author} {\bibfnamefont {N.}~\bibnamefont
  {Urbanek}}, \bibinfo {author} {\bibfnamefont {S.}~\bibnamefont {Toggweiler}},
  \bibinfo {author} {\bibfnamefont {R.}~\bibnamefont {Seelos}}, \bibinfo
  {author} {\bibfnamefont {P.}~\bibnamefont {Jamshidi}}, \bibinfo {author}
  {\bibfnamefont {T.~J.}\ \bibnamefont {Resink}},\ and\ \bibinfo {author}
  {\bibfnamefont {P.}~\bibnamefont {Erne}},\ }\href@noop {} {\bibfield
  {journal} {\bibinfo  {journal} {Atherosclerosis}\ }\textbf {\bibinfo {volume}
  {221}},\ \bibinfo {pages} {124} (\bibinfo {year} {2012})}\BibitemShut
  {NoStop}%
\bibitem [{\citenamefont {Kamiya}\ \emph {et~al.}(1984)\citenamefont {Kamiya},
  \citenamefont {Bukhari},\ and\ \citenamefont {Togawa}}]{kamiya1984adaptive}%
  \BibitemOpen
  \bibfield  {author} {\bibinfo {author} {\bibfnamefont {A.}~\bibnamefont
  {Kamiya}}, \bibinfo {author} {\bibfnamefont {R.}~\bibnamefont {Bukhari}},\
  and\ \bibinfo {author} {\bibfnamefont {T.}~\bibnamefont {Togawa}},\
  }\href@noop {} {\bibfield  {journal} {\bibinfo  {journal} {Bulletin of
  mathematical biology}\ }\textbf {\bibinfo {volume} {46}},\ \bibinfo {pages}
  {127} (\bibinfo {year} {1984})}\BibitemShut {NoStop}%
\bibitem [{\citenamefont {Chen}\ \emph {et~al.}(2012)\citenamefont {Chen},
  \citenamefont {Jiang}, \citenamefont {Li}, \citenamefont {Hu}, \citenamefont
  {Bu}, \citenamefont {Cai},\ and\ \citenamefont {Du}}]{chen2012haemodynamics}%
  \BibitemOpen
  \bibfield  {author} {\bibinfo {author} {\bibfnamefont {Q.}~\bibnamefont
  {Chen}}, \bibinfo {author} {\bibfnamefont {L.}~\bibnamefont {Jiang}},
  \bibinfo {author} {\bibfnamefont {C.}~\bibnamefont {Li}}, \bibinfo {author}
  {\bibfnamefont {D.}~\bibnamefont {Hu}}, \bibinfo {author} {\bibfnamefont
  {J.-w.}\ \bibnamefont {Bu}}, \bibinfo {author} {\bibfnamefont
  {D.}~\bibnamefont {Cai}},\ and\ \bibinfo {author} {\bibfnamefont {J.-l.}\
  \bibnamefont {Du}},\ }\href@noop {} {\bibfield  {journal} {\bibinfo
  {journal} {PLoS Biology}\ }\textbf {\bibinfo {volume} {10}},\ \bibinfo
  {pages} {e1001374} (\bibinfo {year} {2012})}\BibitemShut {NoStop}%
\bibitem [{\citenamefont {Banavar}\ \emph {et~al.}(2000)\citenamefont
  {Banavar}, \citenamefont {Colaiori}, \citenamefont {Flammini}, \citenamefont
  {Maritan},\ and\ \citenamefont {Rinaldo}}]{banavar2000topology}%
  \BibitemOpen
  \bibfield  {author} {\bibinfo {author} {\bibfnamefont {J.~R.}\ \bibnamefont
  {Banavar}}, \bibinfo {author} {\bibfnamefont {F.}~\bibnamefont {Colaiori}},
  \bibinfo {author} {\bibfnamefont {A.}~\bibnamefont {Flammini}}, \bibinfo
  {author} {\bibfnamefont {A.}~\bibnamefont {Maritan}},\ and\ \bibinfo {author}
  {\bibfnamefont {A.}~\bibnamefont {Rinaldo}},\ }\href@noop {} {\bibfield
  {journal} {\bibinfo  {journal} {Physical Review Letters}\ }\textbf {\bibinfo
  {volume} {84}},\ \bibinfo {pages} {4745} (\bibinfo {year}
  {2000})}\BibitemShut {NoStop}%
\bibitem [{\citenamefont {Durand}(2006)}]{durand2006architecture}%
  \BibitemOpen
  \bibfield  {author} {\bibinfo {author} {\bibfnamefont {M.}~\bibnamefont
  {Durand}},\ }\href@noop {} {\bibfield  {journal} {\bibinfo  {journal}
  {Physical Review E—Statistical, Nonlinear, and Soft Matter Physics}\
  }\textbf {\bibinfo {volume} {73}},\ \bibinfo {pages} {016116} (\bibinfo
  {year} {2006})}\BibitemShut {NoStop}%
\bibitem [{\citenamefont {Bohn}\ and\ \citenamefont
  {Magnasco}(2007)}]{bohn2007structure}%
  \BibitemOpen
  \bibfield  {author} {\bibinfo {author} {\bibfnamefont {S.}~\bibnamefont
  {Bohn}}\ and\ \bibinfo {author} {\bibfnamefont {M.~O.}\ \bibnamefont
  {Magnasco}},\ }\href@noop {} {\bibfield  {journal} {\bibinfo  {journal}
  {Physical review letters}\ }\textbf {\bibinfo {volume} {98}},\ \bibinfo
  {pages} {088702} (\bibinfo {year} {2007})}\BibitemShut {NoStop}%
\bibitem [{\citenamefont {Kurz}(2000)}]{kurz2000}%
  \BibitemOpen
  \bibfield  {author} {\bibinfo {author} {\bibfnamefont {H.}~\bibnamefont
  {Kurz}},\ }\href@noop {} {\bibfield  {journal} {\bibinfo  {journal} {J.
  Neuro-Oncol.}\ }\textbf {\bibinfo {volume} {50(1-2)}},\ \bibinfo {pages} {17}
  (\bibinfo {year} {2000})}\BibitemShut {NoStop}%
\bibitem [{\citenamefont {Hu}\ \emph {et~al.}(2012)\citenamefont {Hu},
  \citenamefont {Cai},\ and\ \citenamefont {Rangan}}]{hu2012}%
  \BibitemOpen
  \bibfield  {author} {\bibinfo {author} {\bibfnamefont {D.}~\bibnamefont
  {Hu}}, \bibinfo {author} {\bibfnamefont {D.}~\bibnamefont {Cai}},\ and\
  \bibinfo {author} {\bibfnamefont {A.}~\bibnamefont {Rangan}},\ }\href@noop {}
  {\bibfield  {journal} {\bibinfo  {journal} {PLoS One 7, e45444}\ } (\bibinfo
  {year} {2012})}\BibitemShut {NoStop}%
\bibitem [{\citenamefont {Eichmann}\ \emph {et~al.}(2005)\citenamefont
  {Eichmann}, \citenamefont {Yuan}, \citenamefont {Moyon}, \citenamefont
  {Lenoble}, \citenamefont {Pardanaud},\ and\ \citenamefont
  {Br{\'e}ant}}]{eichmann2005}%
  \BibitemOpen
  \bibfield  {author} {\bibinfo {author} {\bibfnamefont {A.}~\bibnamefont
  {Eichmann}}, \bibinfo {author} {\bibfnamefont {L.}~\bibnamefont {Yuan}},
  \bibinfo {author} {\bibfnamefont {D.}~\bibnamefont {Moyon}}, \bibinfo
  {author} {\bibfnamefont {F.}~\bibnamefont {Lenoble}}, \bibinfo {author}
  {\bibfnamefont {L.}~\bibnamefont {Pardanaud}},\ and\ \bibinfo {author}
  {\bibfnamefont {C.}~\bibnamefont {Br{\'e}ant}},\ }\href@noop {} {\bibfield
  {journal} {\bibinfo  {journal} {Int J Dev Biol}\ }\textbf {\bibinfo {volume}
  {49}},\ \bibinfo {pages} {259} (\bibinfo {year} {2005})}\BibitemShut
  {NoStop}%
\bibitem [{\citenamefont {Scianna}\ \emph {et~al.}(2013)\citenamefont
  {Scianna}, \citenamefont {Bell},\ and\ \citenamefont
  {Preziosi}}]{scianna2013}%
  \BibitemOpen
  \bibfield  {author} {\bibinfo {author} {\bibfnamefont {M.}~\bibnamefont
  {Scianna}}, \bibinfo {author} {\bibfnamefont {C.~G.}\ \bibnamefont {Bell}},\
  and\ \bibinfo {author} {\bibfnamefont {L.}~\bibnamefont {Preziosi}},\
  }\href@noop {} {\bibfield  {journal} {\bibinfo  {journal} {Journal of
  theoretical biology}\ }\textbf {\bibinfo {volume} {333}},\ \bibinfo {pages}
  {174} (\bibinfo {year} {2013})}\BibitemShut {NoStop}%
\bibitem [{\citenamefont {Qi}\ \emph {et~al.}(2024)\citenamefont {Qi},
  \citenamefont {Chang}, \citenamefont {Wang}, \citenamefont {Chen},
  \citenamefont {Baek}, \citenamefont {Hsiai},\ and\ \citenamefont
  {Roper}}]{qi2024}%
  \BibitemOpen
  \bibfield  {author} {\bibinfo {author} {\bibfnamefont {Y.}~\bibnamefont
  {Qi}}, \bibinfo {author} {\bibfnamefont {S.-S.}\ \bibnamefont {Chang}},
  \bibinfo {author} {\bibfnamefont {Y.}~\bibnamefont {Wang}}, \bibinfo {author}
  {\bibfnamefont {C.}~\bibnamefont {Chen}}, \bibinfo {author} {\bibfnamefont
  {K.~I.}\ \bibnamefont {Baek}}, \bibinfo {author} {\bibfnamefont
  {T.}~\bibnamefont {Hsiai}},\ and\ \bibinfo {author} {\bibfnamefont
  {M.}~\bibnamefont {Roper}},\ }\href@noop {} {\bibfield  {journal} {\bibinfo
  {journal} {Proceedings of the National Academy of Sciences}\ }\textbf
  {\bibinfo {volume} {121}},\ \bibinfo {pages} {e2310993121} (\bibinfo {year}
  {2024})}\BibitemShut {NoStop}%
\bibitem [{\citenamefont {Zamir}\ and\ \citenamefont
  {Budwig}(2002)}]{zamir2002}%
  \BibitemOpen
  \bibfield  {author} {\bibinfo {author} {\bibfnamefont {M.}~\bibnamefont
  {Zamir}}\ and\ \bibinfo {author} {\bibfnamefont {R.}~\bibnamefont {Budwig}},\
  }\href@noop {} {\bibfield  {journal} {\bibinfo  {journal} {Appl. Mech. Rev.}\
  }\textbf {\bibinfo {volume} {55}},\ \bibinfo {pages} {B35} (\bibinfo {year}
  {2002})}\BibitemShut {NoStop}%
\bibitem [{\citenamefont {Ang}\ and\ \citenamefont {Jowitt}(2005)}]{ang2005}%
  \BibitemOpen
  \bibfield  {author} {\bibinfo {author} {\bibfnamefont {W.~K.}\ \bibnamefont
  {Ang}}\ and\ \bibinfo {author} {\bibfnamefont {P.~W.}\ \bibnamefont
  {Jowitt}},\ }\href@noop {} {\bibfield  {journal} {\bibinfo  {journal}
  {Engineering Optimization}\ }\textbf {\bibinfo {volume} {37}},\ \bibinfo
  {pages} {277} (\bibinfo {year} {2005})}\BibitemShut {NoStop}%
\bibitem [{\citenamefont {Chatterjee}(2025)}]{chatterjee2025data}%
  \BibitemOpen
  \bibfield  {author} {\bibinfo {author} {\bibfnamefont {P.}~\bibnamefont
  {Chatterjee}},\ }\href@noop {} {\bibinfo {title} {Remodeling simulations}},\
  \bibinfo {howpublished}
  {\url{https://github.com/purbachat/PulsatileAdaptation}} (\bibinfo {year}
  {2025})\BibitemShut {NoStop}%
\bibitem [{\citenamefont {Barnard}\ \emph {et~al.}(1966)\citenamefont
  {Barnard}, \citenamefont {Hunt}, \citenamefont {Timlake},\ and\ \citenamefont
  {Varley}}]{barnard1966}%
  \BibitemOpen
  \bibfield  {author} {\bibinfo {author} {\bibfnamefont {A.}~\bibnamefont
  {Barnard}}, \bibinfo {author} {\bibfnamefont {W.}~\bibnamefont {Hunt}},
  \bibinfo {author} {\bibfnamefont {W.}~\bibnamefont {Timlake}},\ and\ \bibinfo
  {author} {\bibfnamefont {E.}~\bibnamefont {Varley}},\ }\href@noop {}
  {\bibfield  {journal} {\bibinfo  {journal} {Biophysical journal}\ }\textbf
  {\bibinfo {volume} {6}},\ \bibinfo {pages} {717} (\bibinfo {year}
  {1966})}\BibitemShut {NoStop}%
\end{thebibliography}%

\pagebreak

\onecolumngrid
\appendix

\setcounter{equation}{0}
\setcounter{subsection}{0}
\setcounter{figure}{0}
\renewcommand{\theequation}{A\arabic{equation}}
\renewcommand{\thefigure}{A\arabic{figure}}
\renewcommand{\thesubsection}{A\arabic{subsection}}

\section*{Appendix}

\subsection{Single Vessel Dynamics}\label{sec:dynamics}
We consider a fluid with viscosity $i$ and density $\rho$ flowing through a single compliant vessel, treated as a thin cylinder whose length $L$ is much larger than its radius $R$. Assuming rotational symmetry, the local current and pressure in the vessel only depends on the on the axial and radial positions $z$ and $a$. With the axial and radial fluid velocities denoted by $u_z(a,z,t)$ and $u_a(a,z,t)$ respectively, and the fluid pressure by $p(a,z,t)$, the incompressibility condition and the linearized Navier-Stokes equation can be written as
\begin{align}
\vec{\nabla}\cdot\vec{u}&=\frac{\partial u_z}{\partial z}+\frac{1}{a}\frac{\partial}{\partial a}(a u_a)=0,\label{eq:incomp1}\\
\vec{\nabla}p&+\rho\frac{\partial\vec{u}}{\partial t}-\mu\nabla^2\vec{u}=0.\label{eq:NS1}
\end{align}

The axial current $Q(z,t)$, pressure $P(z,t)$ and wall shear stress $\sigma(z,t)$ is defined as
\begin{align}    
    Q(z,t)&=\int dA\> u_z(a,z,t),\label{eq:axQ}\\
    P(z,t)&=A(z,t)^{-1}\int dA\> p(a,z,t),\label{eq:axP}\\
   \sigma(z,t)&=-\mu\frac{\partial u_z(a,z,t)}{\partial a}\bigg|_{a=R},\label{eq:axSig}\\
\end{align}

where the integral in Eq.~\ref{eq:axQ} and Eq.~\ref{eq:axP} is over the cross-sectional area $A(z,t)$ at the axial position $z$ and time $t$. We assume that the vessel
cross sectional area scales linearly with the fluid pressure, but does not vary appreciably with $z$ on the short timescale $t$ of flow relaxation, but only on the longer adaptation timescale $t'$. In other words, $A(z, t) = A_{t'}+cP (z, t)$, where $c$ is the compliance of the vessel, $A_{t'}=\pi R_{t'}^2$ is the cross-sectional area at the adaptation time $t'$, and we make the small deformation approximation $cP (z, t)<<A_{t'}$, such that at short timescales the area of cross-section (and radius) is approximately constant along the entire length of the vessel. From here on we will drop the subscript $t'$ from the constant area (and radius) of the vessel at short timescales, which will evolve on the longer adaptation time-scale through the remodeling rule.  For pulsatile flow with a well defined frequency $\omega$, the axial velocity and axial pressure gradient at a particular position $z=z^*$ can be decomposed into Fourier modes as follows

\begin{align}
 u_z(a,z^*,t)&=\sum_{n=-\infty}^{n=\infty} \tilde{u}_z^{(n)}(a,z^*)e^{\iota n \omega t},\label{vel_decomp}\\
-\nabla_z P(a,z^*,t)&=\sum_{n=-\infty}^{n=\infty} \tilde{G}_z^{(n)}(a,z^*) e^{\iota n \omega t}.\label{grad_pressure_decomp}
\end{align}
The axial component of Eq.~\ref{eq:NS1} at $z=z^*$ can then be written for each Fourier mode as
\begin{equation}
\left(\tilde{u}_z^{(n)}(a,z^*)\right)^{''}+\frac{1}{a}\left(\tilde{u}_z^{(n)}(a,z^*)\right)^{'}-\frac{\iota n\omega\rho}{\mu}\tilde{u}_z^{(n)}(a,z^*)=-\frac{\tilde{G}_z^{(n)}(a,z^*)}{\mu},\label{NS_fourier}
\end{equation}
where $(')$ denotes differentiation with respect to the radial position $a$. Introducing the Womersley number $W_o^{(n)}=R \sqrt{n\omega \rho/\mu}$, the solution of Eq.~\ref{eq:NS1} is
\begin{equation}
    \tilde{u}_z^{(n)}(a,z^*)=\frac{\tilde{G}_z^{(n)}(a,z^*)}{\iota n\omega\rho}\left(1-\frac{J_o\left(\iota^{3/2}W_o^{(n)} a/R\right)}{J_o\left(\iota^{3/2}W_o^{(n)}\right)}\right),\label{eq:vel_sol}
\end{equation}
where $J_0$ is the zeroth order Bessel function of the first kind. The wall shear-stress $\sigma$, the axial current $Q$ and the axial pressure $P$ can also be decomposed into Fourier modes as
\begin{align}
     \sigma(z,t)&=\sum_{n=-\infty}^{n=\infty} \tilde{\sigma}^{(n)}(z)e^{\iota n \omega t},\label{eq:shear_stess_decomp}\\
    Q(z,t)&=\sum_{n=-\infty}^{n=\infty} \tilde{Q}^{(n)}(z) e^{\iota n \omega t},\label{eq:current_decomp}\\
    P(z,t)&=\sum_{n=-\infty}^{n=\infty} \tilde{P}^{(n)}(z) e^{\iota n \omega t}.\label{eq:pressure_decomp}
\end{align}
Using Eq.~\ref{eq:axQ}, Eq.~\ref{eq:axSig} and Eqs.~\ref{eq:vel_sol}-~\ref{eq:current_decomp}, the relation between the wall shear stress and current at $z=z^*$, and therefore for any axial position $z$ can be derived to be
\begin{equation}
    \sigma(z,t)=\sum_{n=-\infty}^{n=\infty} \tilde{f}^{(n)}\tilde{Q}^{(n)}(z)e^{\iota n \omega t},\label{eq:sig_current}\\
\end{equation}
with
\begin{equation}\label{eq:fn}
    \tilde{f}^{(n)}=
    \begin{cases}
      \frac{1}{R^3}, & \text{if}\ n=0 \\
      \frac{\iota n \omega\rho J_1\left(x^{\left(n\right)}\right)}{\pi R\left(x^{\left(n\right)} J_0\left(x^{\left(n\right)}\right)-2J_1\left(x^{\left(n\right)}\right)\right)}, & \text{if}\ n\neq0
    \end{cases}
\end{equation}
where $x^{\left (n\right)}=\iota^{3/2}W_o^{(n)}$, and $\tilde{f}^{(-n)}=(\tilde{f}^{(n)})^{*}$. Note that in the non-pulsatile case with $n=0$, the wall shear stress has a simple relationship to the current given by $\sigma=Q/R^3$. Next, we will derive Eq. $6$ and $7$ of the main text, by integrating Eq.~\ref{eq:incomp1} and Eq.~\ref{eq:NS1} over the area of cross-section. Noting that the expansion rate of the wall must be equal to the radial velocity at the wall, i.e.\
\begin{equation}
    \frac{\partial A}{\partial t}=2\pi R \frac{\partial R}{\partial t}=2\pi R u_a(R,z,t),\label{eq:wall_vel}
\end{equation}
we get by integrating Eq.~\ref{eq:incomp1} over $A$
\begin{align}\label{eq:incomp2}
    0&=\int\>dA \left(\frac{\partial u_z}{\partial z}+\frac{1}{a}\frac{\partial}{\partial a}(au_a)\right),\nonumber\\
    &=\frac{\partial Q}{\partial z}+2\pi R u_a(R,z,t),\nonumber\\
    &=\frac{\partial Q}{\partial z}+\frac{\partial A}{\partial t},\nonumber\\
    \implies0&=\frac{\partial Q}{\partial z}+c\frac{\partial P}{\partial t}.
\end{align}
Assuming that the fluid is Newtonian and
flows in a laminar way, the axial velocity follows the Hagen-Poiseuille equation
\begin{equation}
    u_z(a,z,t)=U(z,t)\left(1-\frac{a^2}{R^2}\right),\label{eq:Poiseuille}
\end{equation}
such that $Q=\int dA\>u_z=\frac{1}{2}AU$ and \begin{equation}
    \frac{\partial u_z}{\partial a}\bigg|_{a=R}=-\frac{2U}{R}=-\frac{4Q}{AR}.\label{eq:uz_vs_r}
\end{equation}
Then, integrating the axial component of Eq.~\ref{eq:NS1} over $A$ and diving by $A$ gives us
\begin{align}\label{eq:NS2}
    0&=\frac{1}{A}\int dA\>\left(\frac{\partial p}{\partial z}+\rho\frac{\partial u_z}{\partial t}-\mu\left(\frac{\partial^2u_z}{\partial z^2}+\frac{1}{a}\frac{\partial}{\partial a}\left(a\frac{\partial u_z}{\partial a}\right)\right)\right),\nonumber\\
    &=\frac{\partial P}{\partial z}+\frac{\rho}{A}\frac{\partial Q}{\partial t}-\frac{\mu}{A}\frac{\partial^2Q}{\partial z^2}-\frac{2\pi\mu}{A}R\frac{\partial u_z}{\partial a}\bigg|_{a=R},\nonumber\\
    &=\frac{\partial P}{\partial z}+\frac{\rho}{A}\frac{\partial Q}{\partial t}+\frac{8\pi\mu}{A^2}\left(Q-\frac{A}{8\pi}\frac{\partial^2Q}{\partial z^2}\right),
\end{align}
where we have used Eq.~\ref{eq:uz_vs_r} in the last step. We will assume that the term $\frac{A}{8\pi}\frac{\partial^2Q}{\partial z^2}$
can be ignored in comparison to $Q$, which holds when the current has approximately a sinusoidal or exponential form in space, with a wavelength or exponential length-scale of decay $\xi$ which is much smaller than the vessel radius $R$. This is because in such scenarios the second derivative of $Q$ with respect to $z$ introduces an inverse dependence on the square of $\xi$, meaning $\frac{A}{8\pi}\frac{\partial^2Q}{\partial z^2}\propto\frac{A}{\xi^2}$, which is negligibly small. Identifying the fluid inertia as $\ell=\rho/A$ and the flow resistance as $r=8\pi\mu/A^2$, Eq.~\ref{eq:incomp2} and Eq.~\ref{eq:NS2} gives us the coupled flow volume and momentum conservation equations (Eq. $6$ and Eq. $7$ in the main text) in terms of the axial current and pressure
\begin{subequations}
\begin{equation}
    \frac{\partial Q}{\partial z}+c\frac{\partial P}{\partial t} = 0,\label{eq:incomp3}
\end{equation}
\begin{equation}
    \frac{\partial P}{\partial z}+\ell\frac{\partial Q}{\partial t}+rQ = 0.\label{eq:NS3}
\end{equation}
\end{subequations}
Of important note is that while $A$ (and $R$) evolves on the timescale of adaptation, $c$, $l$ and $r$ can be considered to be approximately constant on short time scales with the small deformation approximation. This approximation also allows us to express $c$ as
\begin{equation}\label{eq:c}
    c\approx \frac{3A}{2E}\frac{R}{h},
\end{equation}
where $E$ is the Young's modulus of the elastic vessel wall, and $h$ is the wall thickness \cite{barnard1966}. Further assuming that the wall thickness changes proportional to the radius, i.e $R/h$ remains approximately constant over the time scale of adaptation, the compliance $c$ scales proportional to $A$, whereas $\ell\propto A^{-1}$ and $r\propto A^{-2}$. From $c$, $l$ and $r$, we can derive a characteristic length scale $\lambda=(2/r)\sqrt{l/c}$, a characteristic time scale $\tau=2l/r$ and a characteristic admittance $\alpha=\sqrt{c/l}$, all three of which scale proportional to $A=\pi R^2$. These derived parameters can be used to recast the flow and momentum conservation equations in a more symmetric form, 
\begin{subequations}
\begin{equation}
    \lambda\frac{\partial Q}{\partial z}+\alpha\tau\frac{\partial P}{\partial t} = 0,\label{eq:incomp3}
\end{equation}
\begin{equation}
    \alpha\lambda\frac{\partial P}{\partial z}+\ell\frac{\partial Q}{\partial t}+rQ = 0,\label{eq:NS3}
\end{equation}
\end{subequations}
and the parameters themselves can be expressed as a constant times the area of cross-section as
\begin{subequations}
\begin{equation}    \lambda=\lambda_0\left(R/R_0\right)^2,\label{eq:lam}
\end{equation}
\begin{equation}
    \tau=\tau_0\left(R/R_0\right)^2,\label{eq:tau}
\end{equation}
\begin{equation}
    \alpha=\alpha_0\left(R/R_0\right)^2,\label{eq:alpha}
\end{equation}
\end{subequations}
where $R_0$ is a typical radius for the compliant vessel. As elaborated on in the main text, $\lambda$ characterizes the length scale over which pulsatility is damped in the vessel, whereas $\tau$ sets the short timescale of the problem, capturing the time taken for the flow to relax locally. Eqs.~\ref{eq:incomp3} and \ref{eq:NS3} have exact solutions as long as some combinantion of two boundary conditions are known. Here, we will consider the case in which the pressures at the $z=0$ and $z=L$ ends of the vessel are known, though it is worth noting that solutions can be obtained for any combination of two non-redundant boundary conditions, be they current, pressure, or mixed. We get
\begin{subequations}
\begin{equation}
\tilde{Q}^{\left(n\right)}\left(z\right) = \frac{\iota n\omega\tau\alpha}{k\left(n\omega\tau\right)}\frac{\tilde{P}^{\left(n\right)}\left(0\right)\cosh\left(\frac{L-z}{\lambda}k\left(n\omega\tau\right)\right)-\tilde{P}^{\left(n\right)}\left(L\right)\cosh\left(\frac{z}{\lambda}k\left(n\omega\tau\right)\right)}{\sinh\left(\frac{L}{\lambda}k\left(n\omega\tau\right)\right)},
\label{eq:Qsol}
\end{equation}

\begin{equation}
\tilde{P}^{\left(n\right)}\left(z\right) = \frac{\tilde{P}^{\left(n\right)}\left(0\right)\sinh\left(\frac{L-z}{\lambda}k\left(n\omega\tau\right)\right)+\tilde{P}^{\left(n\right)}\left(L\right)\sinh\left(\frac{z}{\lambda}k\left(n\omega\tau\right)\right)}{\sinh\left(\frac{L}{\lambda}k\left(n\omega\tau\right)\right)},
\label{eq:Psol}
\end{equation}
\end{subequations}

\noindent where
\begin{equation}
k\left(n\omega\tau\right) = \sqrt{\iota n\omega\tau\left(2+\iota n\omega\tau\right)}.
\label{eq:kdef}
\end{equation}
Eqs.~\ref{eq:Qsol} and \ref{eq:Psol} together with Eqs.~\ref{eq:current_decomp}-\ref{eq:pressure_decomp}, constitute the solutions for the axial current and pressure supported by the flow dynamics model represnted by Eqs.~\ref{eq:incomp3}-\ref{eq:NS3}.  

\subsection{Single Vessel Dissipation}\label{sec:diss}

The rate at which energy has to be expended to overcome the resistance to flow presented by the vessel wall, i.e. the dissipation per unit time, is given by the product of the total force of resistance and the
average flow velocity. Since the wall shear stress $\sigma(z,t)$ is the frictional force per unit area exerted by the fluid on the wall at the axial position $z$, the total frictional force of resistance is simply the product of the shear stress and the surface area of the vessel wall, i.e. $F_{fric}=\sigma(z,t)\cdot 2\pi R L$. The average flow velocity is given by $v = Q(z, t)/(\pi R^2)$. Thus, the total rate of energy dissipation per unit length of the vessel is given by \cite{zamir2002}
\begin{equation}\label{eq:diss_def}
    D(z,t)\equiv F_{fric}\cdot v=\frac{2}{R}\sigma(z,t)Q(z,t).
\end{equation}
We will use $D=\langle\sigma Q\rangle$, averaged over vessel length $L$ and one period of oscillation $T=2\pi/\omega$, as the adaptation drive seen by each individual elastic vessel during remodeling. Utilizing Eq.~\ref{eq:sig_current} along with Eq.~\ref{eq:Qsol}, we can solve for single vessel adaptation drive as

\begin{align}
\left\langle\sigma Q\right\rangle &= \frac{1}{LT}\int_{0}^{L}dz\int_{0}^{T}dt\>\sigma\left(z,t\right) Q\left(z,t\right),\nonumber\\
&=\frac{1}{LT} \int_{0}^{L}dz\int_{0}^{T}dt\>\sum_{n,n'}\tilde{f}^{(n)}\tilde{Q}^{\left(n\right)}\left(z\right)\tilde{Q}^{\left(n'\right)}\left(z\right)e^{\iota\omega\left(n+n'\right)t} = \frac{1}{L}\int_{0}^{L}dz\sum_{n}\tilde{f}^{(n)}\abs{\tilde{Q}^{\left(n\right)}\left(z\right)}^{2} \nonumber\\
&= \frac{1}{L}\sum_{n=-\infty}^{n=\infty}\tilde{f}^{(n)}\int_{0}^{L}dz\abs{\frac{\iota n\omega \tau\alpha}{k\left(n\omega \tau\right)}\frac{\tilde{P}^{\left(n\right)}\left(0\right)\cosh\left(\frac{L-z}{\lambda}k\left(n\omega \tau\right)\right)-\tilde{P}^{\left(n\right)}\left(L\right)\cosh\left(\frac{z}{\lambda}k\left(n\omega \tau\right)\right)}{\sinh\left(\frac{L}{\lambda}k\left(n\omega \tau\right)\right)}}^{2} \nonumber\\
&= \sum_{n=-\infty}^{n=\infty}\frac{\lambda}{L}\tilde{f}^{(n)}\abs{\frac{\iota n\omega \tau\alpha }{k\left(n\omega \tau\right)\sinh\left(\frac{L}{\lambda}k\left(n\omega \tau\right)\right)}}^{2} \nonumber\\
&\quad\quad\quad \cdot\left(\left(\abs{\tilde{P}^{\left(n\right)}\left(0\right)}^{2}+\abs{\tilde{P}^{\left(n\right)}\left(L\right)}^{2}\right)\frac{\text{Im}\left(k\left(n\omega \tau\right)\sinh\left(\frac{L}{\lambda}k\left(n\omega \tau\right)\right)\cosh\left(\frac{L}{\lambda}k^{*}\left(n\omega \tau\right)\right)\right)}{\text{Im}\left(\left(k\left(n\omega \tau\right)\right)^{2}\right)}\right. \nonumber\\
&\quad\quad\quad \left.-\left(\tilde{P}^{\left(n\right)}\left(0\right)\tilde{P}^{\left(n\right)*}\left(L\right)+\tilde{P}^{\left(n\right)*}\left(0\right)\tilde{P}^{\left(n\right)}\left(L\right)\right)\frac{\text{Im}\left(k\left(n\omega \tau\right)\sinh\left(\frac{L}{\lambda}k\left(n\omega \tau\right)\right)\right)}{\text{Im}\left(\left(k\left(n\omega \tau\right)\right)^{2}\right)}\right).
\label{eq:sigQave}
\end{align}

Simplifying the various terms of Eq. \ref{eq:sigQave} by recognizing that $(k(y))^{2}=2iy-y^{2}$, and freely bringing select real terms inside the imaginary operators, we get

\begin{align}
\left\langle\sigma Q\right\rangle &= \sum_{n=-\infty}^{n=\infty}\frac{\alpha\lambda}{2L}n\omega \tau\alpha\tilde{f}^{(n)}\left(\left(\abs{\tilde{P}^{\left(n\right)}\left(0\right)}^{2}+\abs{\tilde{P}^{\left(n\right)}\left(L\right)}^{2}\right)\text{Im}\left(\frac{k\left(n\omega \tau\right)\sinh\left(\frac{L}{\lambda}k\left(n\omega \tau\right)\right)\cosh\left(\frac{L}{\lambda}k^{*}\left(n\omega \tau\right)\right)}{\abs{k\left(n\omega \tau\right)\sinh\left(\frac{L}{\lambda}k\left(n\omega \tau\right)\right)}^{2}}\right)\right. \nonumber\\
&\quad\quad\quad \left.-2\text{Re}\left(\tilde{P}^{\left(n\right)}\left(0\right)\tilde{P}^{\left(n\right)*}\left(L\right)\right)\text{Im}\left(\frac{k\left(n\omega \tau\right)\sinh\left(\frac{L}{\lambda}k\left(n\omega \tau\right)\right)}{\abs{k\left(n\omega \tau\right)\sinh\left(\frac{L}{\lambda}k\left(n\omega \tau\right)\right)}^{2}}\right)\right) \nonumber\\
&= \sum_{n=-\infty}^{n=\infty}\frac{\alpha\lambda}{2L}n\omega \tau\alpha\tilde{f}^{(n)}\left(\left(\abs{\tilde{P}^{\left(n\right)}\left(0\right)}^{2}+\abs{\tilde{P}^{\left(n\right)}\left(L\right)}^{2}\right)\text{Im}\left(\left(\frac{\cosh\left(\frac{L}{\lambda}k\left(n\omega \tau\right)\right)}{k\left(n\omega \tau\right)\sinh\left(\frac{L}{\lambda}k\left(n\omega \tau\right)\right)}\right)^{*}\right)\right. \nonumber\\
&\quad\quad\quad \left.-2\text{Re}\left(\tilde{P}^{\left(n\right)}\left(0\right)\tilde{P}^{\left(n\right)*}\left(L\right)\right)\text{Im}\left(\left(\frac{1}{k\left(n\omega \tau\right)\sinh\left(\frac{L}{\lambda}k\left(n\omega \tau\right)\right)}\right)^{*}\right)\right).
\label{eq:sigQave_simp1}
\end{align}

\noindent Finally, we transform imaginary operators into real operators by noting that $\text{Im}(z^{*})=\text{Re}(\iota z)$. We can also freely bring in the factor of $n\omega \tau\alpha$ to transform Eq. \ref{eq:sigQave_simp1} into

\begin{align}
\left\langle\sigma Q\right\rangle &= \sum_{n=-\infty}^{n=\infty}\frac{\alpha\lambda}{2L}\tilde{f}^{(n)}\left(\left(\abs{\tilde{P}^{\left(n\right)}\left(0\right)}^{2}+\abs{\tilde{P}^{\left(n\right)}\left(L\right)}^{2}\right)\text{Re}\left(\frac{\iota n\omega \tau\alpha}{k\left(n\omega \tau\right)}\frac{\cosh\left(\frac{L}{\lambda}k\left(n\omega \tau\right)\right)}{\sinh\left(\frac{L}{\lambda}k\left(n\omega \tau\right)\right)}\right)\right. \nonumber\\
&\quad\quad\quad \left.-2\text{Re}\left(\tilde{P}^{\left(n\right)}\left(0\right)\tilde{P}^{\left(n\right)*}\left(L\right)\right)\text{Re}\left(\frac{\iota n\omega \tau\alpha}{k\left(n\omega \tau\right)}\frac{1}{\sinh\left(\frac{L}{\lambda}k\left(n\omega \tau\right)\right)}\right)\right),\nonumber\\
\implies \left\langle\sigma Q\right\rangle &=\left(\frac{\alpha\lambda}{2L}\right)^2\text{Re}\left(\tilde{f}^{(0)}\right) \left(\tilde{P}^{(0)}\left(0\right)-\tilde{P}^{(0)}\left(L\right)\right)^2\nonumber\\
&\quad\quad\quad +\sum_{n=1}^{n=\infty} \left(\frac{\alpha\lambda}{L}\right)\text{Re}\left(\tilde{f}^{(n)}\right)\left(\abs{\tilde{P}^{(n)}\left(0\right)}^{2}+\abs{\tilde{P}^{(n)}\left(L\right)}^{2}\right)\text{Re}\left(\frac{\iota n\omega\tau\alpha}{k(n\omega\tau)}\coth\left(\frac{L}{\lambda}k(n\omega\tau)\right)\right) \nonumber\\
&\quad\quad\quad -2\sum_{n=1}^{n=\infty}\left(\frac{\alpha\lambda}{L}\right)\text{Re}\left(\tilde{f}^{(n)}\right)\text{Re}\left(\tilde{P}^{(n)}\left(0\right)\left(\tilde{P}^{(n)}\left(L\right)\right)^{*}\right)\text{Re}\left(\frac{\iota n\omega\tau\alpha}{k(n\omega\tau)}\csch\left(\frac{L}{\lambda}k(n\omega\tau)\right)\right).
\label{eq:sigQave_simp2}
\end{align}

\noindent Note that for a single vessel, the expression for $\left\langle\sigma Q\right\rangle$ differs from the expression for $\left\langle Q^2\right\rangle$ used as the adaptation drive in the widely used steady-flow remodeling rule (Eq. $1$ in the main text) only by the modification of each frequency mode by $\text{Re}\left(\tilde{f}^{(n)}\right)$ given in Eq.~\ref{eq:fn}. For periodically driven networks, the pulsatile-flow remodeling rule with $\left\langle\sigma Q\right\rangle$ as the adapation drive (Eq. $5$ in the main text) can be expected to generate a more biologically realistic response. 
\subsection{Network Construction and Solution}\label{sec:Laplacian}

We now consider a network of nodes connected by compliant vessels. Let $Q_{ij}(z,t)$ and $P_{i,j}(z,t)$ be the current and pressure in the vessel that connects node $i$ to node $j$. The position variable, $z$, is measured such that $z=0$ denotes the location at node $i$ and $z=L_{ij}$ denotes the location at node $j$. Importantly, under interchange of node indices the pressure is a scalar quantity and satisfies $P_{ij}(z,t)=P_{ji}(L_{i,j}-z,t)$ while the current is a psuedoscalar quantity and satisfies $Q_{ij}(z,t)=-Q_{ji}(L_{ij}-z,t)$. $L_{ij}$ is of course a parameter and invariant to such interchange, as are the vessel parameters $\lambda_{ij}$, $\tau_{ij}$, and $\alpha_{ij}$. Enforcing continuity of pressure across the network allows us to define the pressure at node $i$ as

\begin{equation}
P_{i}\left(t\right) = P_{ij}\left(0,t\right) \quad\forall\quad j\in\mathcal{N}_{i},
\label{eq:prescont}
\end{equation}

\noindent where $\mathcal{N}_{i}$ is the set of all nodes connected to node $i$ by a single vessel. Moreover, conservation of mass requires that $H_{i}(t)$, which is the total external current input at node $i$, must equal the total current leaving the node, which can be expressed as

\begin{equation}
H_{i}\left(t\right) = \sum_{j\in\mathcal{N}_{i}}Q_{ij}\left(0,t\right).
\label{eq:masscon}
\end{equation}

Assuming that the network is driven by a pulsatile source with base frequency $\omega$ allows us to expand Eq. \ref{eq:masscon} into its Fourier modes and combine it with Eqs. \ref{eq:Qsol} and \ref{eq:prescont} to produce

\begin{align}
\tilde{H}_{i}^{\left(n\right)} &= \sum_{j\in\mathcal{N}_{i}}\tilde{Q}_{ij}^{\left(n\right)}\left(0\right) = \sum_{j\in\mathcal{N}_{i}}\left.\frac{\iota n\omega_{0}\tau_{ij}\alpha_{ij}}{k\left(n\omega_{0}\tau_{ij}\right)}\frac{\tilde{P}_{ij}^{\left(n\right)}\left(0\right)\cosh\left(\frac{L_{ij}-z}{\lambda_{ij}}k\left(n\omega_{0}\tau_{ij}\right)\right)-\tilde{P}_{ij}^{\left(n\right)}\left(L_{ij}\right)\cosh\left(\frac{z}{\lambda_{ij}}k\left(n\omega_{0}\tau_{ij}\right)\right)}{\sinh\left(\frac{L_{ij}}{\lambda_{ij}}k\left(n\omega_{0}\tau_{ij}\right)\right)}\right|_{z=0} \nonumber\\
&= \sum_{j\in\mathcal{N}_{i}}\frac{\iota n\omega_{0}\tau_{ij}\alpha_{ij}}{k\left(n\omega_{0}\tau_{ij}\right)}\frac{\tilde{P}_{i}^{\left(n\right)}\cosh\left(\frac{L_{ij}}{\lambda_{ij}}k\left(n\omega_{0}\tau_{ij}\right)\right)-\tilde{P}_{j}^{\left(n\right)}}{\sinh\left(\frac{L_{ij}}{\lambda_{ij}}k\left(n\omega_{0}\tau_{ij}\right)\right)} = \sum_{j}\mathcal{L}_{ij}^{\left(n\right)}\tilde{P}_{j}^{\left(n\right)},
\label{massconmat}
\end{align}

\noindent where the summation in the final expression is over all $j$, not just nodes in $\mathcal{N}_{i}$, and the network Laplacian matrix is given by

\begin{equation}
\mathcal{L}_{ij}^{\left(n\right)} = \begin{cases}
\sum_{\gamma\in\mathcal{N}_{i}}\alpha_{i\gamma}\frac{\iota n\omega_{0}\tau_{i\gamma}}{k\left(n\omega_{0}\tau_{i\gamma}\right)}\frac{\cosh\left(\frac{L_{i\gamma}}{\lambda_{i\gamma}}k\left(n\omega_{0}\tau_{i\gamma}\right)\right)}{\sinh\left(\frac{L_{i\gamma}}{\lambda_{i\gamma}}k\left(n\omega_{0}\tau_{i\gamma}\right)\right)} & j=i \\
-\alpha_{ij}\frac{\iota n\omega_{0}\tau_{ij}}{k\left(n\omega_{0}\tau_{ij}\right)}\frac{1}{\sinh\left(\frac{L_{ij}}{\lambda_{ij}}k\left(n\omega_{0}\tau_{ij}\right)\right)} & j\in\mathcal{N}_{i} \\
0 & \text{otherwise} \end{cases},
\label{Lapdef}
\end{equation}

\noindent which for the non-pulsatile mode $n=0$ limits to

\begin{equation}
\mathcal{L}_{ij}^{\left(0\right)} = \begin{cases}
\sum_{\gamma\in\mathcal{N}_{i}}\alpha_{i\gamma}\frac{\lambda_{i\gamma}}{2L_{i\gamma}} & j=i \\
-\alpha_{ij}\frac{\lambda_{ij}}{2L_{ij}} & j\in\mathcal{N}_{i} \\
0 & \text{otherwise} \end{cases}.
\label{Lap0def}
\end{equation}

\noindent Eq. \ref{massconmat} provides a relation between the boundary currents and pressures. Thus, if there are $N$ nodes in the network, then Eq. \ref{massconmat} represents $N$ equations that allow for all $2N$ boundary conditions ($N$ from the boundary currents and $N$ from the node pressures) that can be solved for so long as at least $N$ are known. If the boundary currents are known, the pressure at each node can be evaluated by inverting the network Laplacian as

\begin{equation}
\tilde{P}_i^{\left(n\right)} = \left(\left(\mathcal{L}^{\left(n\right)}\right)^{-1}\tilde{H}^{\left(n\right)}\right)_i .
\label{eq:Linvert}
\end{equation}

\noindent Eq.~\ref{eq:sigQave_simp2} can be written for every vessel in the network. For $N_E$ vessels, each vessel connecting a node $i=1,2,...N$ to a node $j=1,2,...N$ can be identified by the subscript $e=1,2,...N_E$, with the pressures at the two ends of the vessel (i.e. at nodes $i$ and $j$) represented as $P_{in}$ and $P_{out}$ respectively. Denoting the length and values of derived parameters for the $e^{th}$ vessel by $L_e$, $\tau_e$, $\lambda_e$ and $\alpha_e$, Eq.~\ref{eq:sigQave_simp2} can then be written as
\begin{align} 
\left\langle\sigma_e Q_e\right\rangle &=\left(\frac{\alpha_e\lambda_e}{2L_e}\right)^2\text{Re}\left(\tilde{f}_e^{(0)}\right) \left(\tilde{P}_{in}^{(0)}-\tilde{P}_{out}^{(0)}\right)^2\nonumber\\
&\quad\quad\quad +\sum_{n=1}^{n=\infty} \left(\frac{\alpha_e\lambda_e}{L_e}\right)\text{Re}\left(\tilde{f}_e^{(n)}\right)\left(\abs{\tilde{P}_{in}^{(n)}}^{2}+\abs{\tilde{P}_{out}^{(n)}}^{2}\right)\text{Re}\left(\frac{\iota n\omega\tau_e\alpha_e}{k(n\omega\tau_e)}\coth\left(\frac{L_e}{\lambda_e}k(n\omega\tau_e)\right)\right) \nonumber\\
&\quad\quad\quad -2\sum_{n=1}^{n=\infty}\left(\frac{\alpha_e\lambda_e}{L_e}\right)\text{Re}\left(\tilde{f}_e^{(n)}\right)\text{Re}\left(\tilde{P}_{in}^{(n)}\left(\tilde{P}_{out}^{(n)}\right)^{*}\right)\text{Re}\left(\frac{\iota n\omega\tau_e\alpha_e}{k(n\omega\tau_e)}\csch\left(\frac{L_e}{\lambda_e}k(n\omega\tau_e)\right)\right),\nonumber\\
&=\sum_{n=-\infty}^{n=\infty}Re[\tilde{f}_e^{(n)}]\langle Q_e^2\rangle^{(n)},
\label{eq:sigQave_e} 
\end{align}
\noindent where $\langle Q_e^2\rangle^{(n)}$ is the contribution of the $n^{th}$ frequency mode to the mean-squared current $\langle Q_e^2\rangle$. Eq.~\ref{eq:sigQave_e}, together with the solutions for node pressures obtained from Eq.~\ref{eq:Linvert} can thus be used to calculate the driving signal for adaptation in each vessel of the network.

\subsection{Comparison Of Remodeling Rules}\label{sec:rules}
While the remodeling rules considered in the main text (Eq. $1$ and Eq. $5$) are based on the optimization of a well-defined energy functional (Eq. $2$ and Eq.$3$ in the main text), in general it cannot be guaranteed that biological networks have feasible mechanisms to directly drive them towards an optimum architecture that exactly minimizes energy dissipation and metabolic or material costs. On the other hand, the construction of a remodeling rule that is not based on gradient descent down a cost function, involves arbitrary choices regarding what physical quantities (e.g. shear stress, flow rate etc.) or combinations of physical quantities can be locally sensed in the adapting vessels, thereby generating the driving signal for remodeling. All we know with certainty from experimental evidence is that radii evolution must explicitly depend on the wall shear stress, but the vessel averaged wall shear stress by itself has also been shown to be insufficient for driving the growth of vascular networks. 
\begin{figure*}
\begin{center}
\includegraphics[scale=0.9]{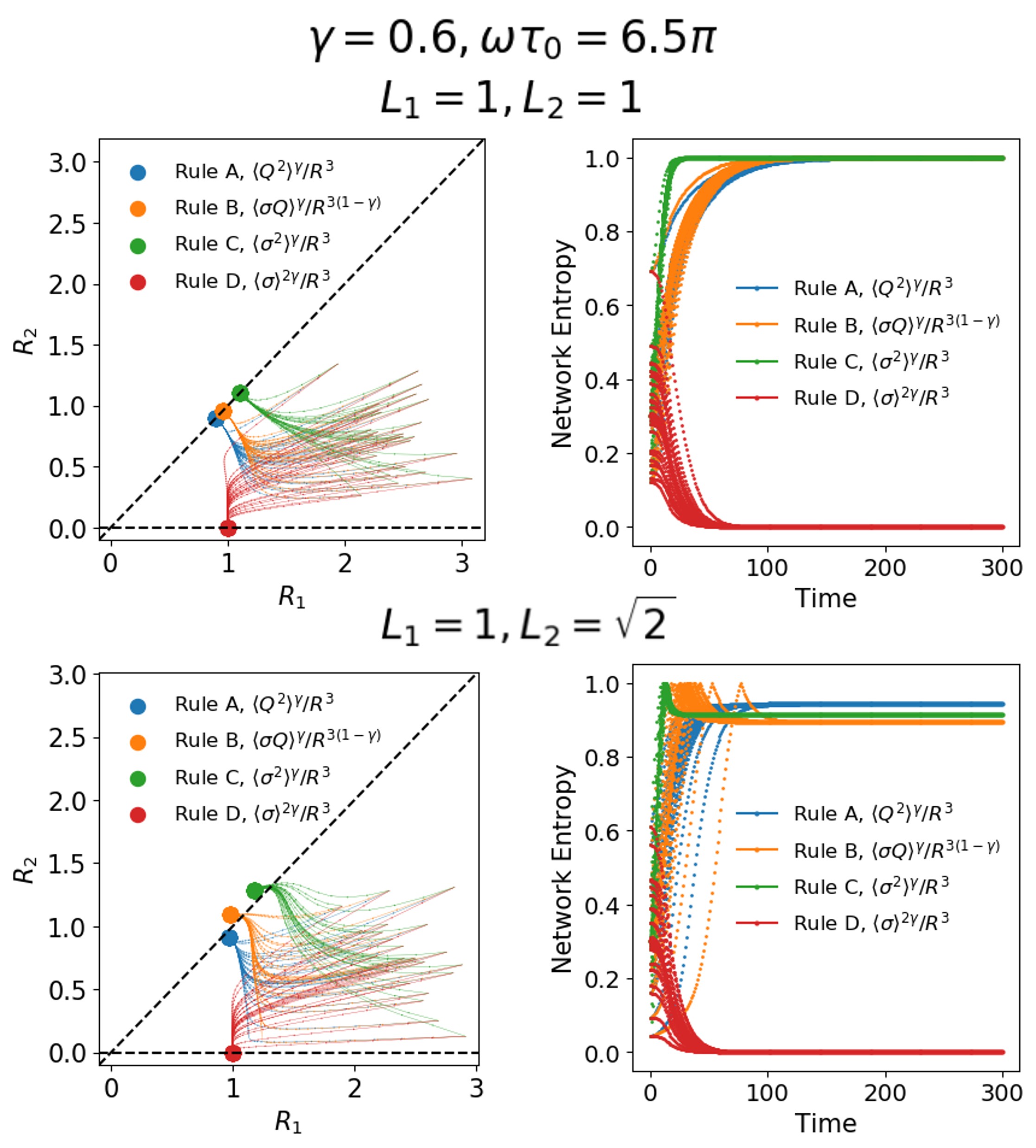}
\end{center}
\caption {\small {Comparison of different adaptation rules used to evolve the two vessel network in Fig. $2$(a) of the main text, with $\gamma=0.55$ and $\omega\tau_0=6.5 \pi$. Top: Equal vessel lengths, $L_1=L_2=1$. The left panel shows the flow diagrams in $R_1-R_2$ space obtained with each rule for the same $25$ initial sets of radii. The right panel shows the time evolution of the network entropy for each rule, for the same $25$ trajectories. Bottom: Unequal vessel lengths, $L_1=1, L_2=\sqrt{2}$. Again, the left panel shows the flow diagrams, and the right panel shows the time evolution of the network entropy with each rule, for $25$ initial sets of radii. In both the equal and unequal vessel length cases, rules $A$, $B$ and $C$, which retain the effect of the pulsatile component of the flow in their driving signals, generate qualitatively identical steady-state structures: a symmetric loop in the equal vessel length case and an asymmetric loop in the unequal vessel length case. The rule $C$, in which the pulsatile contribution averages to zero in the driving signal, leads to shunting or loop destabilization in both the equal and unequal vessel length cases. This confirms that it is indeed the short-term pulsatility, and not the exact mathematical form of the adaptation rule, that leads to robust loop stabilization for $\gamma>1/2$. $R_0=\tau_0=a=b=Q_s=Q_p=1$ and $\lambda_0=2$.}}
    \label{fig:rules}
\end{figure*}

While identifying a `correct' or realistic adaptation rule is difficult for the reasons outlined above, it is useful to first investigate how different the long-term outcomes of different remodeling rules are. If the outcomes of closely related rules (whether gradient descent based or not) lead to qualitatively similar outcomes, we can conclude that the exact functional form of the rule is not as important to the dynamics of adaptation in the presence of pulsatile flows. In this section, we will compare four different remodeling rules, given by
\begin{align}
    Rule\  A:\hspace{10pt}\frac{dR_{e}}{dt'}&=\frac{a\langle Q_e^2\rangle^\gamma}{R_{e}^3}-bR_{e},\label{eq:ruleA}\\ 
    Rule\  B:\hspace{10pt}\frac{dR_{e}}{dt'}&=\frac{a\langle \sigma_e Q_e\rangle^\gamma}{R_{e}^{3(1-\gamma)}}-bR_{e},\label{eq:ruleA}\\ 
    Rule\  C:\hspace{10pt}\frac{dR_{e}}{dt'}&=\frac{a\langle \sigma_e^2\rangle^\gamma}{R_{e}^3}-bR_{e},\label{eq:ruleA}\\ 
    Rule\  D:\hspace{10pt}\frac{dR_{e}}{dt'}&=\frac{a\langle \sigma_e \rangle^{2\gamma}}{R_{e}^3}-bR_{e},\label{eq:ruleA}.
\end{align}
Here, $A$ and $B$ are the gradient descent based rules considered in the main text, whereas $C$ and $D$ are not derived from any energy functional. We use each of these rules to evolve the two vessel network illustrated in Fig. $2$(a) of the main text, for both equal and unequal vessel length cases. Fig.~\ref{fig:rules} shows the results with each rule for the same $25$ sets of initial radii for the two vessels, and for $\gamma=0.6$ and $\omega\tau_0=6.5\pi$. In both the equal vessel length case (top row, $L_1=L_2=1$) and the unequal vessel length case (bottom row, $L_1=1, L_2=\sqrt{2}$), we find that rules $A, B$ and $C$ generate qualitavely identical networks (symmetric and asymmetric loops respectively) and quantitatively similar network entropies as a function of time. Rule $D$ however generates starkly different results than the other three rules, leading to vessel shunting, i.e. loop destabilization in both the equal and unequal vessel length cases. Thus, the three rules $A, B$ and $C$, which retain the pulsatile contribution to the driving signal, lead to effectively similar adaptation outcomes even though one of them ($C$) is not derived from an energy functional. In contrast, rule $C$, where the pulsatile contribution to the driving signal averages to zero, yields fundamentally different long-term structures than the other three rules. This makes us conclude that loop stabilization is less contingent on the exact mathematical form of the driving signal and more so on the ability of the driving signal to retain the pulsatile contributions from the short-term flow dynamics in vessels. 

\subsection{Effect Of Varying The Pulsatile Amplitude}\label{sec:qp}
\begin{figure*}
\begin{center}
\includegraphics[scale=0.657]{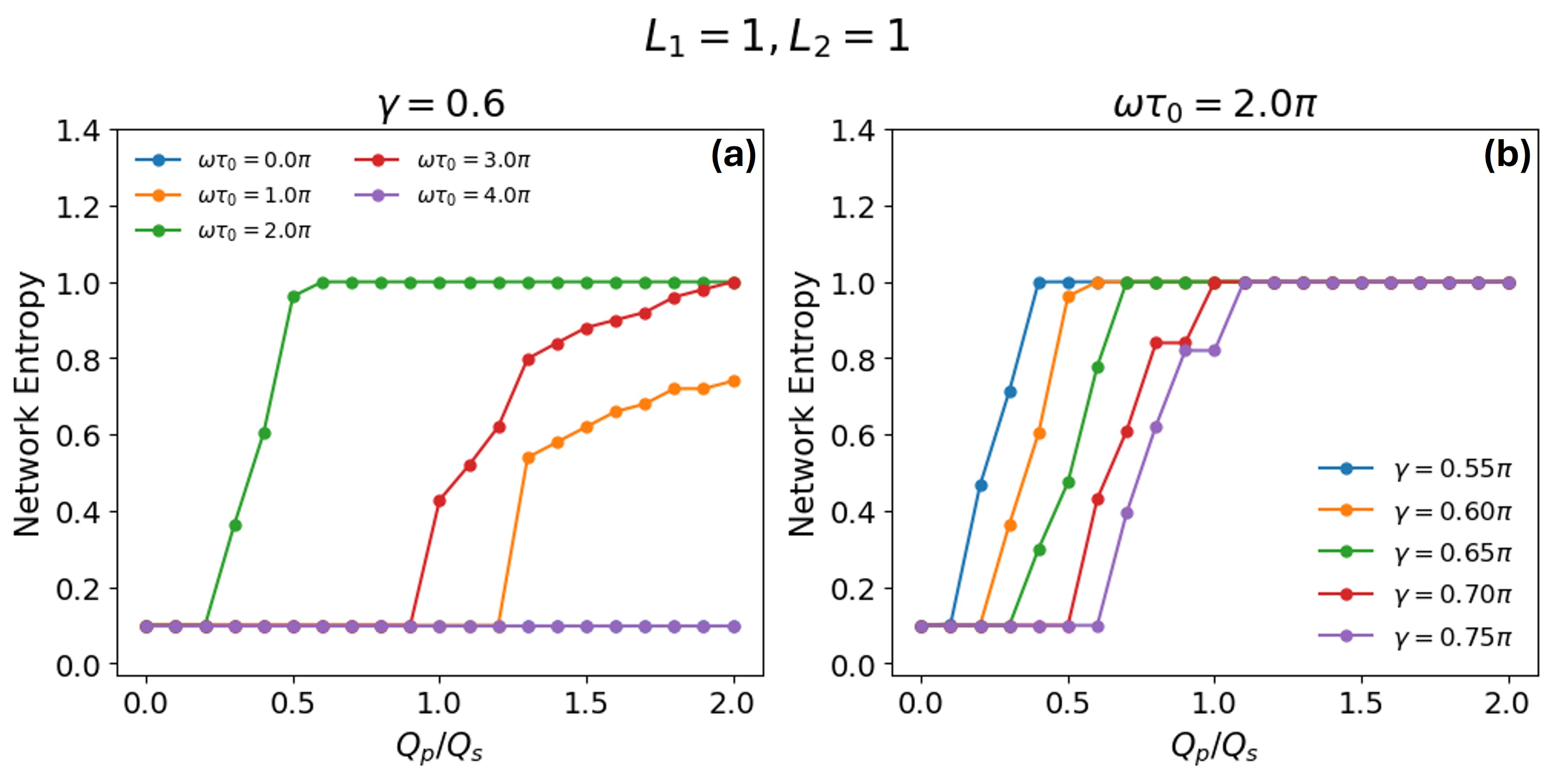}
\end{center}
\caption {\small {Variation of the average network entropy with the ratio of the pulsatile and steady flow amplitudes, $Q_p/Q_s$ of the source current, for the two vessel network depicted in Fig. $2$(a), and for $L_1=L_2=1$. The average is over the same set of initial radii used to generate Fig. $3$(a-f) in the main text. (a) shows results for $\gamma=0.6>1/2$ and different values of $\omega\tau_0$, and (b) shows results for $\omega\tau_0=2\pi$ and different values of $\gamma$. For all $\omega\tau_0$ values in (a) and all $\gamma$ values in (b), the network entropy increases with increasing $Q_p/Q_s$, indicating that larger amplitudes of pulsatility lead to larger ranges of driving frequencies and metabolic cost parameters where loops can be stabilized. $R_0=\tau_0=a=b=1$ and $\lambda_0=2$.}}
    \label{fig:qp}
\end{figure*}

In all results presented in the main text, the steady and pulsatile components of the source current are set to have equal amplitudes ($Q_s=Q_p=1$). Here we investigate the effect of varying $Q_p$ relative to $Q_s$ for the two vessel network depicted in Fig. $2$(a) of the main text. Fig.~\ref{fig:qp} shows how the average network entropy changes with $Q_p/Q_s$ for $L_1=L_2=1$, and for $100$ sets of initial radii. The adaptation rule used is the pulsatile remodeling rule (Eq. $5$). Fig.~\ref{fig:qp}(a) shows the results for different $\omega\tau_0$ values and for a fixed $\gamma=0.6$, and Fig.~\ref{fig:qp}(b) shows the results for different $\gamma$ values and for a fixed $\omega\tau_0=2\pi$. In both Fig.~\ref{fig:qp}(a) and (b), the network entropy increases with increasing $Q_p/Q_s$ for all values of $\omega\tau_0$ and $\gamma$ respectively, indicating that the resonance peaks in the structure phase diagram in $\gamma-\omega\tau_0$ phase space become taller and broader as the pulsatile amplitude increases. This further supports our main result, that the short term pulsatile dynamics of elastic vessels has the dominant impact on the long-term structures  of adapting elastic networks, making loop stabilization more robust when the pulsatile amplitude is larger.
\subsection{Comparison Of Networks With And Without Current Sinks }\label{sec:sinks}
\begin{figure*}
\begin{center}
\includegraphics[scale=0.6]{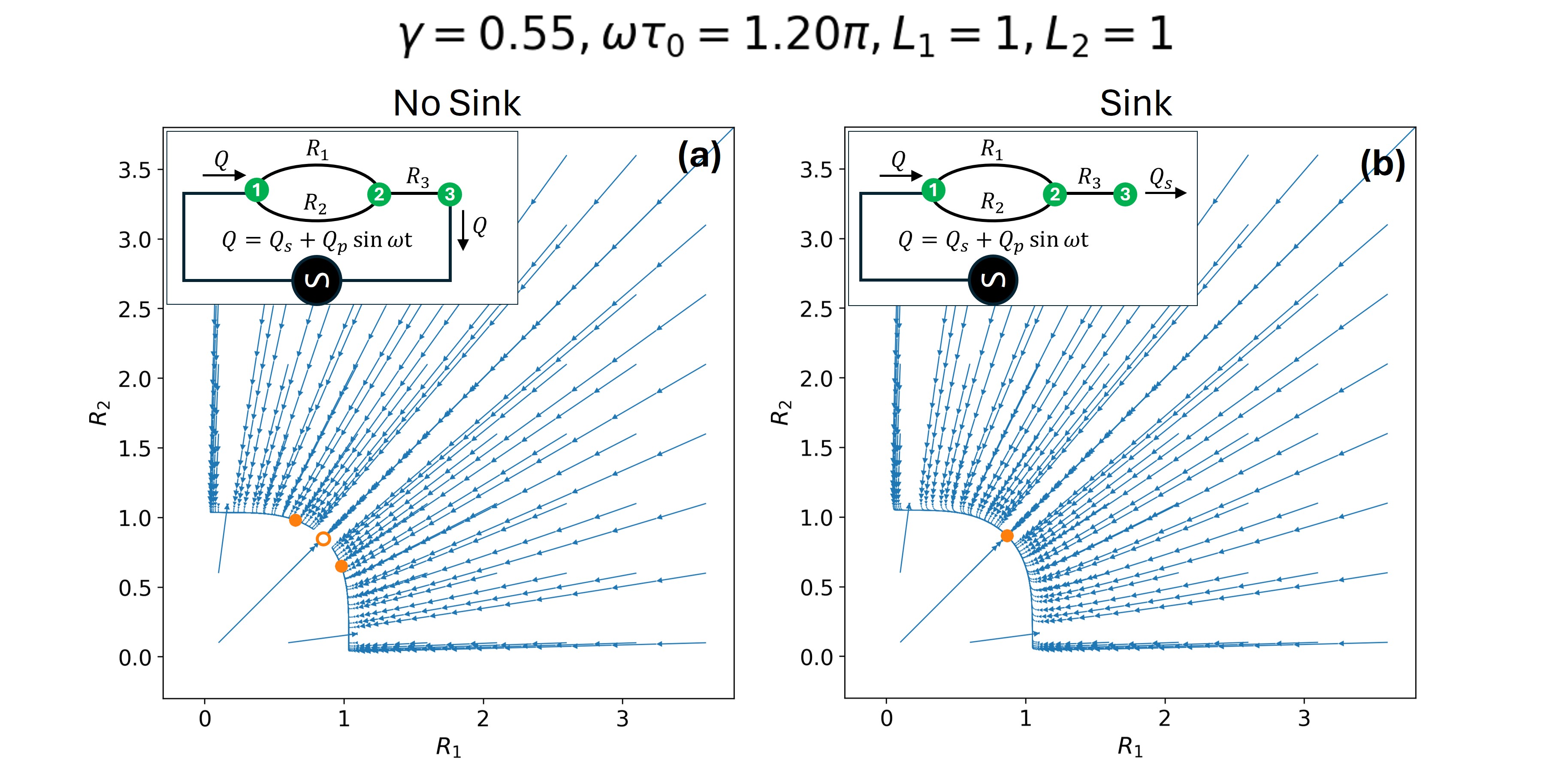}
\end{center}
\caption {\small {Comparison between three vessel networks with and without a current sink, for $\gamma=0.55>1/2, \omega\tau_0=1.2\pi$ and $L_1=L_2=1$ . (a) Flow diagram in the $R_1-R_2$ space in the `No sink' case, where the three vessel network and the pulsatile source comprise a closed circuit (inset). Here $R_1$ and $R_2$ are the radii of the two vessels comprising the loop, and the initial radii of the third vessel $R_3$ is set to $1$ for simplicity. For most initial values of $R_1$ and $R_2$, the long-term steady state is an asymmetric loop. For exactly equal starting values of $R_1$ and $R_2$ a symmetric loop can be stabilized. (a) Flow diagram in the $R_1-R_2$ space in the case with a current sink (node $3$), from which the steady part of the source current can flow out. Once again, $R_3$ is set to have an initial value equal to $1$, and for all initial values of $R_1$ and $R_2$, the long-term steady state is an symmetric loop. $R_0=\tau_0=a=b=Q_s=Q_p=1$ and $\lambda_0=2$.}}
    \label{fig:sinks}
\end{figure*}

In the main text, we only consider toy networks with a pulsatile source current $Q=Q_s+Q_p \sin{(\omega t)}$ entering and leaving through two externally driven nodes, i.e. forming a closed circuit (see Figs. $2$(a), $4$(a) and $6$(a) in the main text). To test the sensitivity of our results to different boundary conditions, we will now consider the case where the current entering the loop is pulsatile, but the current exiting is steady. This is inspired by the arterial portion of the circulatory system that delivers blood to the organs, where pulsatility is greatly attenuated. Specifically in this section we compare the results generated for two different networks with three adaptable vessels and a single loop, evolved through the pulsatile remodeling rule (Eq. $5$). The insets of Fig.~\ref{fig:sinks}(a) and (b) illustrate how the circuits are connected in the two cases. The first case, which we call the `No Sink' case, similar to the two-vessel network considered in the main text. However, it is different from the two vessel case because in addition to the two vessels forming the loop (with radii $R_1$ and $R_2$), there is a third vessel with adaptable radius $R_3$. The same pulsatile current $Q=Q_s+Q_p \sin{\omega t}$ enters through node $1$ and leaves through node $3$, forming the closed circuit. The second network we consider is the `Sink' case, in which the current that enters through node $1$ is the full pulsatile source current $Q$, whereas the current that leaves through node $3$ is only the steady part of $Q$, i.e. $Q_s$. Thus, in the `Sink' case the pulsatile component of the source current is allowed to dissipate in the extra vessel as its radius $R_3$ evolves over time.

Fig.~\ref{fig:sinks} shows the results in each case for $\gamma=0.55>1/2$, $\omega\tau_0=1.2\pi$ and equal vessel lengths $L_1=L_2=1$. For simplicity we always set the initial radii of the third vessel (not involved in the loop) to be $R_3=1$. Fig.~\ref{fig:sinks}(a) and (b) show the flow diagrams for the no-sink and sink cases in the $R_1-R_2$ space, i.e the space of the two vessels comprising the loop.  We find that for all initial values of $R_1$ and $R_2$, the long-term steady state is a symmetric loop in the case with a current sink, while most initial conditions lead to an asymmetric loop ($R_1\neq R_2$) at steady state in the case without sinks. Thus, even with current sinks, the short-term pulsatile dynamics of elastic vessels can stabilize looped architechtures.

\end{document}